\documentclass[prc,aps,groupedaddress,superscriptaddress,twocolumn,nofootinbib,showpacs,showkeys,floatfix,a4paper,10pt]{revtex4-1}

\usepackage{graphicx,colordvi,rotating}
\usepackage{multirow,color}
\usepackage{amsmath,amssymb}
\usepackage[mathscr]{euscript}
\usepackage{cancel}
\usepackage{braket}
\usepackage{cases}
\usepackage{color}
\usepackage{mathtools}
\usepackage{natbib}
\usepackage{graphicx}
\usepackage{dcolumn}
\usepackage{bm}
\usepackage{natbib}
 \usepackage{enumerate}

\usepackage[utf8]{inputenc}
\usepackage{epstopdf}

\begin{document}

\title{Description of the multidimensional potential energy surface in fission of $^{252}$Cf and $^{258}$No}

\author{A. Zdeb}
\email{anna.zdeb@cea.fr}
\affiliation{CEA, DAM, DIF, F-91297 Arpajon, France}
\affiliation{Université Paris-Saclay, CEA, LMCE, F-91680 Bruyères-le-Châtel, France }

\author{M. Warda}
\email{michal.warda@umcs.pl}
\affiliation{Instytut Fizyki, Uniwersytet Marii Curie--Sk\l odowskiej, PL-20031 Lublin, Poland}

\author{L. M. Robledo}
\email{luis.robledo@uam.es}
\affiliation{Departamento de F\'\i sica Te\'orica and CIAFF, Universidad Aut\'onoma de Madrid, Madrid, Spain}
\affiliation{Center for Computational Simulation,
Universidad Polit\'ecnica de Madrid,
Campus de Montegancedo, Bohadilla del Monte, E-28660 Madrid, Spain
}

\date{\today}

\begin{abstract}
The microscopic studies on nuclear fission require the evaluation of 
the potential energy surface as a function of the collective 
coordinates. A reasonable choice of constraints on multipole moments 
should be made to describe the topography of the surface completely 
within a reasonable amount of computing time. We present a detailed 
analysis of fission barriers in the self-consistent 
Hartree-Fock-Bogoliubov approach with the D1S parametrization of the 
Gogny nucleon-nucleon interaction. Two heavy isotopes representing 
different spontaneous fission modes - $^{252}$Cf (asymmetric) and 
$^{258}$No (bimodal) - have been chosen for the analysis. We have shown 
the existence of complicated structures on the energy surface that can 
not be fully described in two-dimensional calculations. We analyze 
apparent problems that can be encountered in this type of calculations: 
multiple solutions for given constraints and transitions between 
various potential energy surfaces. We present possible solutions on how 
to deal with these issues.
\end{abstract}


\maketitle

\section{Introduction}

The accurate description of collective nuclear motion from the ground 
state up to the scission point represents a crucial step to understand the fission 
of the atomic nucleus. An essential ingredient is the behavior of the
binding energy of the nucleus as a function of the parameters characterizing 
the collective degrees of freedom. On its way from the initial ground state configuration
to scission, the nucleus has to tunnel through a 
potential energy barrier determining the time scale of fission half-lives.
Not only the height but also the width and the 
shape of the fission barrier are important for spontaneous fission 
half-lives estimations, while the topography of the potential energy 
surface (PES) is essential to determine fission dynamics. A detailed 
description of the theory of fission and challenges in this field can 
be found in the review papers: Refs. 
\cite{krappe2012,Schunck2016,10.1088/1361-6471/abab4f}.

The first historical attempt to explain and describe the fission 
process was based on the semi-empirical liquid drop model 
assumptions~\cite{bohr} where the interplay between Coulomb repulsion 
and surface energy of a charged and deformed drop of nuclear matter is 
responsible for the splitting~\cite{meitner}. The success of this 
simple approach suggested that collective variables (in an intrinsic 
frame of the nucleus), associated with the shape evolution of a 
fissioning system, are a crucial ingredient in any microscopic analysis 
of fission process. The adiabatic approximation - based on the 
assumption that the time-scale of nucleons motion is much shorter than 
the time-scale of shape changes - allows obtaining a consistent 
description by considering just a small set of collective degrees of 
freedom well adapted to the shape evolution involved in fission.

Many theoretical papers describing the PES in heavy actinides have been 
published recently improving our knowledge of fission of heavy and 
super-heavy nuclei 
\cite{Pashkevich2009,Moller2009,Mirea2010,Afanasjev2010,Kowal2012,Lu2014,Guzman2014,%
pomorski2015,Guzman2016,schmitt2017,jachimowicz2017,pomorski2018,%
Chai_2018,jachimowicz2020,jachimowicz2020a, guzman2020}. In those articles, one can 
distinguish two different approaches to describe the shape of the 
nucleus in the fission process. Both are based on the minimization of 
the energy along the different paths connecting the starting 
configuration (usually the ground state) with the scission 
configurations. In the first approach, which is mostly used with 
macroscopic-microscopic theories, a pre-defined class of nuclear shapes 
defined uniquely in terms of a given set of deformation parameters is 
used to define the set of accessible configurations to be used in the 
minimization of the energy. The number of parameters defines the 
dimensionality of the problem. An increasing number of dimensions 
increases the variety of the shapes used and, therefore, usually 
improves the calculation's quality by providing lower energy solutions. 
This flexibility comes at the expense of calculating the energy for a 
huge number of points in the deformation space. On the other hand, the 
approach benefits from full control over the shapes characterizing the 
evolution of a nucleus from the ground state to scission. The 
alternative approach is mostly used by the microscopic self-consistent 
methods. In the self-consistent procedure, the wave functions along the 
fission path are determined by the minimization of the energy of the 
nucleus within a given set of constraints and assumed symmetries. The 
constraints only provide general guidelines for the nuclear density 
distribution and do not fix its shape uniquely. For a given set of 
constraints, the nuclear shape is free to take any form among the 
allowed ones as to minimize the energy of the nucleus. The advantage of 
this procedure is that we do not have to worry if our space of 
deformation parameters is large enough to describe the shape adopted by 
the nucleus. Nevertheless, the results provided with this method are 
not always unique~\cite{DUBRAY2012} as one can easily land for a given 
set of constraints in a local minimum instead of the absolute one. As a 
consequence, the self-consistent minimization procedure can produce 
several solutions with the same value of the constraints but different 
energies depending, for instance, on the starting wave function used in 
the solution of the variational equations. Those ``multivalued" 
solutions are connected in a higher dimensional deformation space by a 
barrier separating them. A typical example, often seen in axially 
symmetric calculation constraining in the quadrupole moment, is the 
existence of two solutions with the same quadrupole moment but 
different values of the octupole moment in the region of the second 
barrier. The two coexisting minima are linked by a path going through 
the octupole moment in the parameter space. One of the unwanted 
consequences of this kind of situation is the possibility to jump 
between configurations when considering the behavior of quantities as a 
function of the current shape parameter. The leap is unphysical because 
one is skipping the path connecting the coexisting minima. When 
describing this situation, one talks about jumping from one fission 
valley to another. Whether the dynamics of fission justify this 
transition or it is just an artifact due to the limited number of 
constraints used to describe the fission path is a delicate issue that 
will be the subject of the present paper.   


If the constraints are partially released, several different 
solutions may exist as local minima of the potential energy. In this case, one has to decide  
which of them should be considered in the analysis of fission.  
We may even encounter a situation where on the edge between two valleys, a {\it 
fake} or {\it missing} saddle can be found~\cite{DUBRAY2012}. The first 
case occurs when two distinct nuclear shape configurations have got the 
same energy at some coordinate, and the fission paths are incorrectly 
linked together. The missing one means that minimizing the energy with 
a single constraint may pass over a saddle leading to another valley. 
The consequence of a wrong interpretation of the PES may have a 
non-negligible impact on the description of fission dynamics. The barrier heights may be 
improperly evaluated, and the fission valleys may be omitted or 
redundant. 

It has been previously shown that non-geometrical constraints like the 
one associated with pairing correlations may strongly influence the 
fission dynamics scenario because of the strong dependence of the 
collective inertias on the inverse of the pairing gap 
\cite{urin,moretto,baran,PhysRevC.90.054311,PhysRevC.98.034308,schunk}. 
As a consequence, the inclusion of the pairing degree of freedom can 
substantially modify the dynamical evolution of a collective wave 
packet. Moreover, this kind of constraint can play a central role if 
the least action principle is used instead of the least energy one to 
determine the fission path. The treatment of spontaneous fission using 
the least action principle and including the pairing degree of freedom 
leads to a substantial reduction of the spontaneous fission life 
time~\cite{PhysRevC.90.054311,PhysRevC.98.034308,schunk,baran}. 
However, and despite its relevance, we are not considering the pairing 
degree of freedom in the present study. 

This paper is devoted to the discussion of the choice of the multipole 
moments to be used as constraints for the calculation of the PES 
leading to fission in the self-consistent methods. The primary purpose 
is to determine which subspace of collective coordinates in a 
deformation space is sufficient for an accurate description of static 
properties of fissioning nuclei. We will analyze the mechanism of 
creating multiple solutions on the PES in the limited dimensionality of 
the space. The consequences of this behavior will be discussed. We will 
indicate regions where the discontinuity of the PES may be found.

To carry out our investigation, we have chosen two isotopes 
representing two different fission modes - asymmetric 
($^{252}$Cf)~\cite{cfexp,PhysRevC.90.064611,PhysRevC.23.2100} and 
bimodal ($^{258}$No)~\cite{huletPRL, hulet}. With this choice, we cover 
a broad and representative range of fission modes in heavy nuclei. The 
total half-life of the radioactive isotope $^{252}$Cf is 
$t_{1/2}=2.645(8)$ y with alpha decay as the dominant branch. Only 
3.092(8)\% of the fissioning nuclides undergo spontaneous fission. On 
the other hand, $^{258}$No has a far shorter half live $t_{1/2}=1.2(2)$ 
ms and fission is the principal decay channel \cite{audi20170301}.


\section{Theoretical framework \label{THEORY}}

The present study has been carried out using the self-consistent 
constrained Hartree-Fock-Bogoliubov (HFB) method with the finite-range 
density-dependent nucleon-nucleon Gogny interaction. We use the 
well-reputed parametrization D1S \cite{Berger90} that has been 
extensively used in the literature to describe many different nuclear 
structure phenomena \cite{Peru2014,Robledo_2018} including the microscopic description
of fission \cite{BERGER1991365,War02,DELAROCHE2006103,BARAN2015442,wardaPhysRevC.86.014322}.

In the HFB method \cite{Schunck2016,Robledo_2018}, the nuclear states are 
obtained as the solutions of the HFB equation, which is derived by requiring 
that the mean value of the routhian is a minimum: 
\begin{equation}
\delta(\left \langle{\Phi}\right | \widehat{H}-\lambda_Z \widehat{Z}- \lambda_N \widehat{N}-\sum\limits_{ij} \lambda_{ij} \widehat{Q}_{ij}\left|{\Phi}\right>)=0.
\end{equation}
Here $\widehat{H}$ is the microscopic Hamiltonian, $\lambda_N$ and 
$\lambda_Z$ are the Lagrange multipliers used to fix the number of neutrons 
$N$ and protons $Z$, while  $\lambda_{ij}$ are the Lagrange multipliers 
associated with the average value of the multipole moments $\hat{Q}_{ij}$ with multipolarity
$i$ and projection $j$. In this work 
the quadrupole $(Q_{20})$, octupole $(Q_{30})$, hexadecapole $(Q_{40})$ and 
triaxial quadrupole ($Q_{22}$) deformation parameters are considered. The 
equation is solved by expanding the creation and annihilation quasiparticle
operators of the Bogoliubov transformation in a 
harmonic oscillator basis with oscillator length parameters optimized for each set 
of collective deformation parameters as to minimize the binding energy. Most of the results are obtained in the 
axial regime with an axially symmetric deformed oscillator basis with $N_\perp=15$ and 
$N_z=22$. This basis is well suited to describe elongated shapes along the 
$z$ axis as those typical of fission. Beyond mean-field, two-body kinetic energy correction and 
rotational energy correction are included in the calculation of the binding energy. 
In the calculations with non-zero triaxial multipole moment $Q_{22}$ 
reflection symmetry is preserved, and therefore odd multipole moments are 
zero by construction and not considered in the discussion. For the ``triaxial" 
calculations an oscillator basis containing $N=18$ shells is used. 

The multipole moment operators are defined as
\begin{equation}
\hat Q_{lm}=\frac{1}{\sqrt{2(1+\delta_{m0})}}(\hat M_{lm}+r_m(-1)^m\hat M_{l-m})
\end{equation}
with $r_m=1$ if $m\geq 0$ and -1 if $m < 0$. The raw multipole operators $\hat M_{lm}$
are given by
\begin{eqnarray}
\hat M_{lm}&=&\sqrt{\frac{4\pi}{2l+1}} r^l Y_{lm}(\theta,\varphi)\\
&=& \sqrt{\frac{(l-m)!}{(l+m)!}} r^l P_{lm}[cos(\theta)]e^{im\phi} \;,\nonumber
\end{eqnarray}
where $Y_{lm}$ are 
spherical harmonics and $P_{lm}$ are associated Legendre polynomials. 
Using the standard definition of the spherical
harmonics we obtain $\hat Q_{20}=z^2-\frac{1}{2}(x^2+y^2)$, 
$\hat Q_{30}=z^3-\frac{3}{2}(x^2+y^2)z $, 
$\hat Q_{40}= z^4-3(x^2+y^2)z^2+\frac{3}{8}(x^2+y^2)^2$ and $\hat Q_{22}=\sqrt{3}/2(x^2-y^2)$.

The computations were performed using the self-consistent HFB solver {\it HFBaxial}
that uses the approximate second order gradient method
\cite{PhysRevC.84.014312} to solve the HFB equations and the formulas of 
\cite{EGIDO199713} for an accurate  evaluation of the matrix elements of the Gaussian central potential 
for the large basis required in fission.
The program starts from an initial wave function and iteratively minimizes
the energy subject to a given set of constraints. In this procedure, the shape of the 
nucleus is adjusted to fulfill the conditions imposed by the constraint 
parameters. Usually, the initial configuration is taken from a 
neighbor, previously computed, wave function. As the initial configuration
is expected to be ``close" to the sought solution this choice represents
a way to decrease the number of iterations required in the minimization process and
therefore, represents substantial computer time savings.
The final solution may depend on the starting point in specific cases. 
This feature will be thoroughly discussed below.

As it has been mentioned in the Introduction, the use of a limited set of 
constraints always introduces some uncertainties in which wave functions
belong to the neighborhood of which local minima. To help in the 
characterization of the different solutions of the HFB equation 
it is convenient to use a technique 
based on searching for  discontinuities in the matter density distribution 
all over the PES. The so-called {\it density distance} 
$D_{\rho\rho'}$~\cite{DUBRAY2012} has been used to detect such 
discontinuities in calculations considering
the pair of shape parameters $Q_{20}-Q_{30}$ 
as well as $Q_{30}-Q_{40}$. For given configurations with matter density
distributions $\rho$ and $\rho'$, the density distance is defined as:
\begin{equation}
D_{\rho\rho'}= \int|\rho({\bf r})-\rho'({\bf r})| d \bf r\;.
\label{dens_dist}
\end{equation}
Density distance remains small when nuclear 
shapes are similar, and it increases wherever there is a substantial 
change of nuclear shape for the two densities considered.


\section{Results}
 
\subsection{The PES in many coordinates \label{subsPES} }

\begin{figure}

\includegraphics[width=0.7\columnwidth, angle=270]{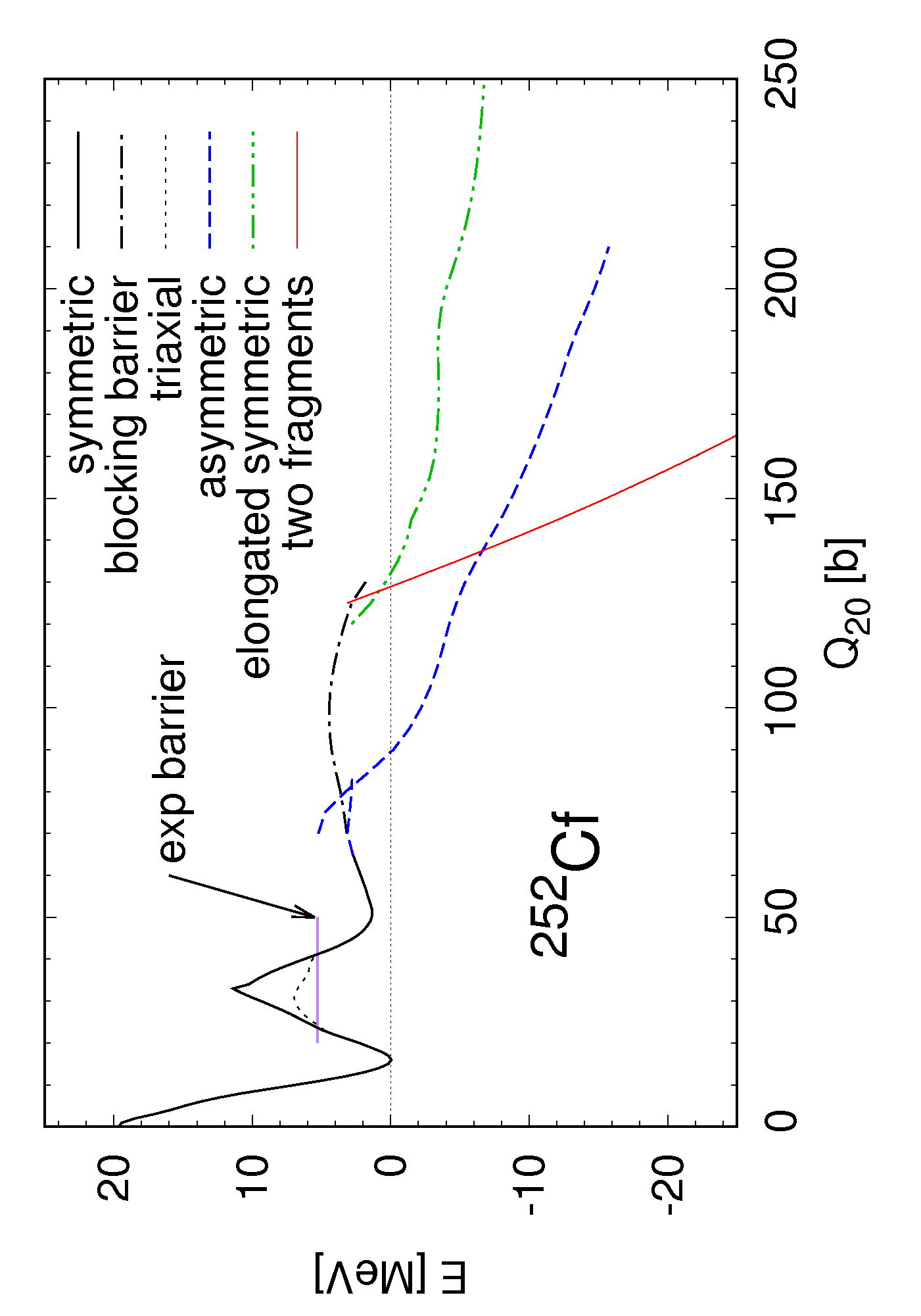}\\
\includegraphics[width=0.7\columnwidth, angle=270]{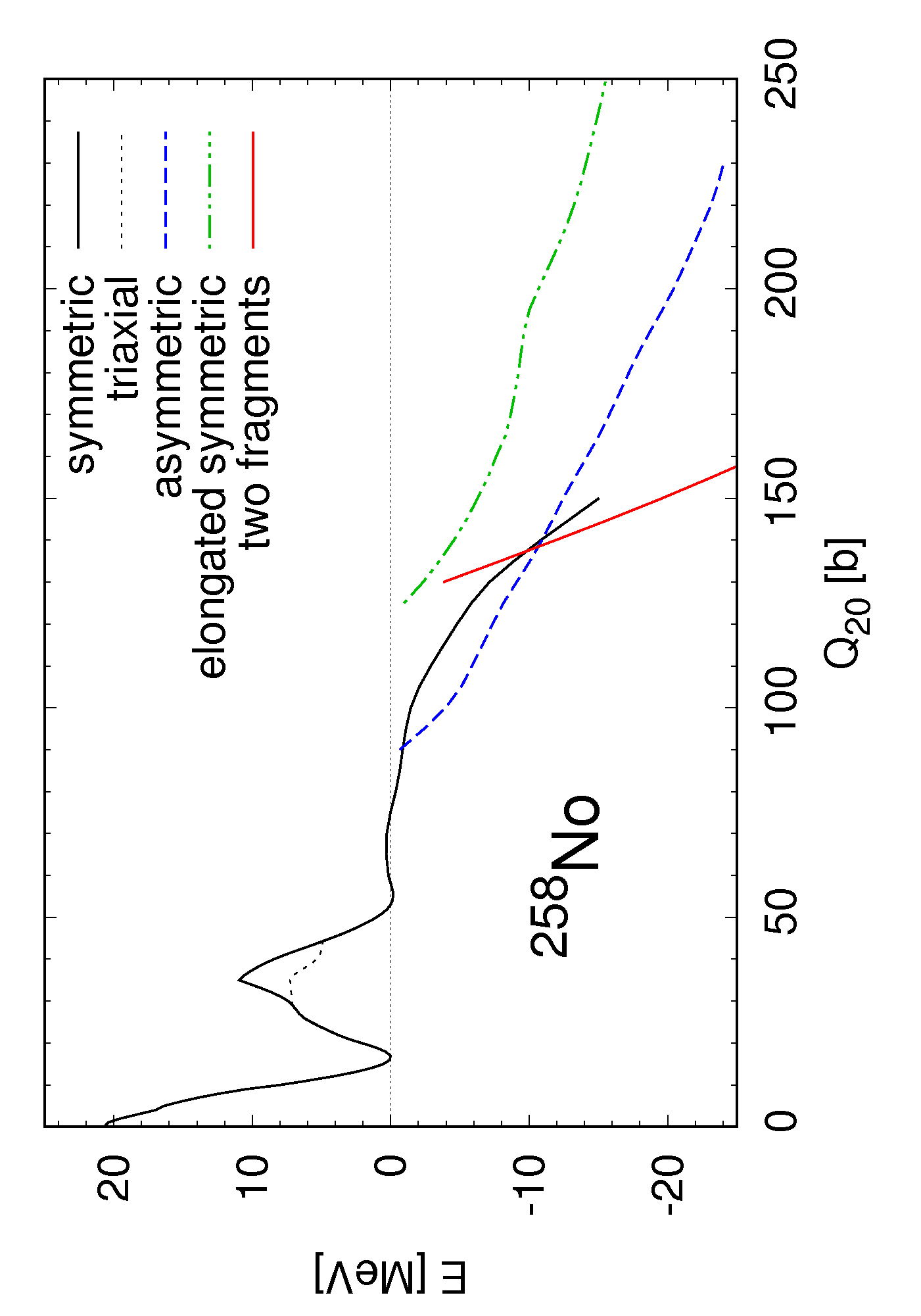}
\caption{Various fission paths for $^{252}$Cf (top) and $^{258}$No (bottom) 
are plotted as a function of $Q_{20}$.  See text for details.\label{PATHcfno}}
\end{figure}

\begin{figure}
\includegraphics[width=0.7\columnwidth, angle=270]{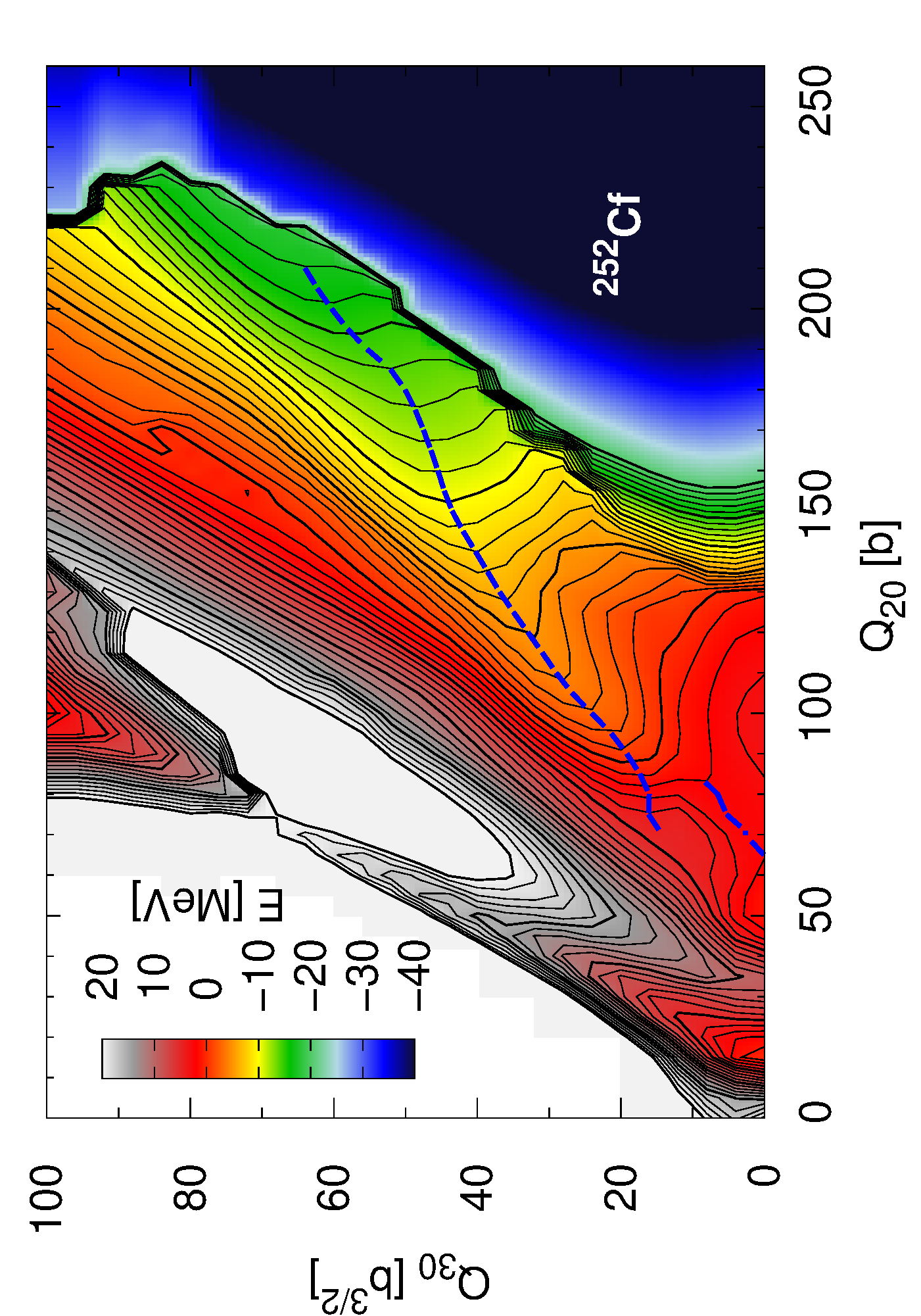}\\
\includegraphics[width=0.7\columnwidth, angle=270]{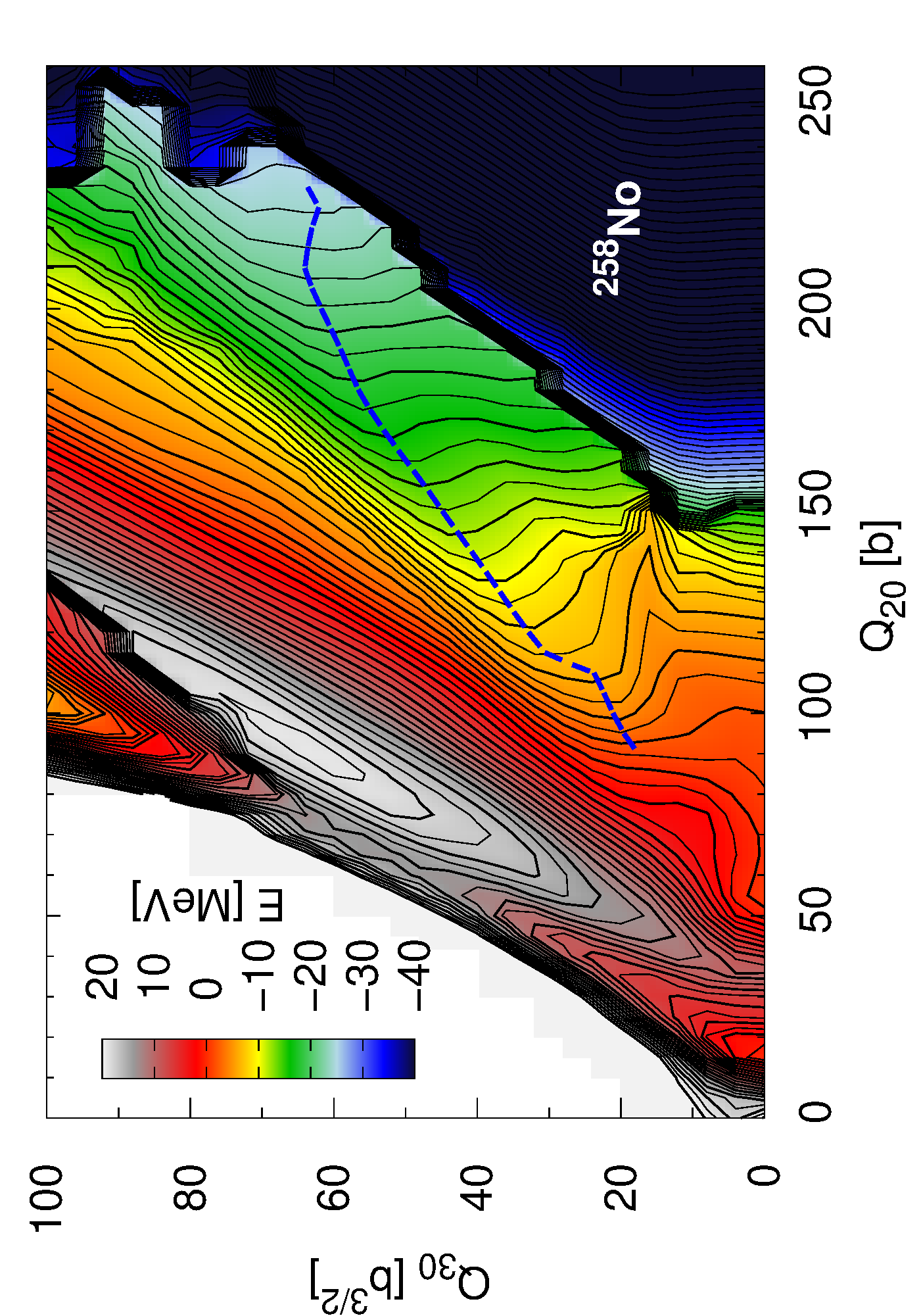}
\caption{The PES of $^{252}$Cf (top) and$^{258}$No (bottom) as a function of 
$Q_{20}$ and $Q_{30}$ are shown as contour and color plots. 
Constant energy contours are plotted every 1~MeV. 
Asymmetric fission path is represented by a blue dashed line. 
\label{PEScfno}}
\end{figure}

\begin{figure}
\includegraphics[width=0.7\columnwidth, angle=270]{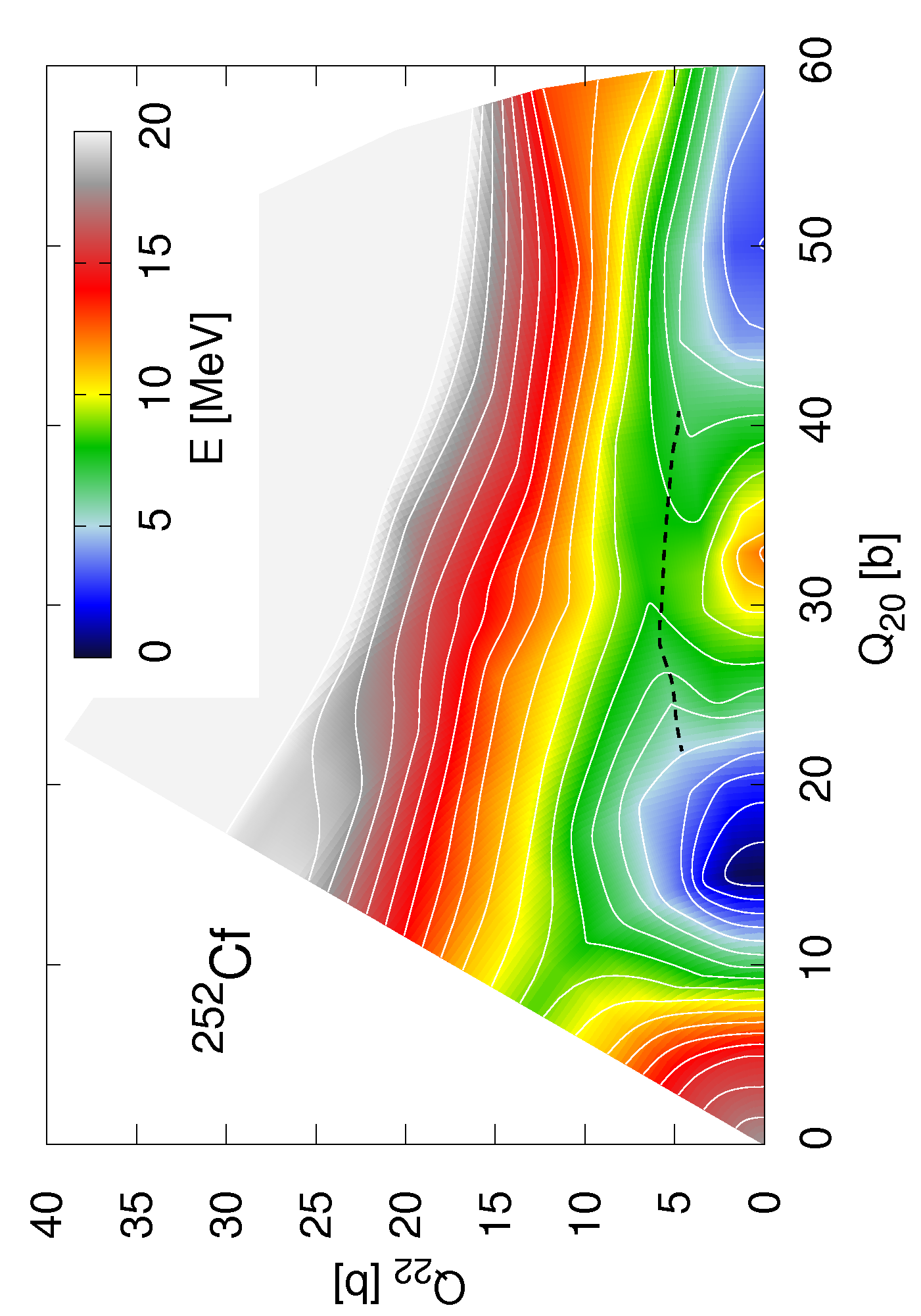}\\
\includegraphics[width=0.7\columnwidth, angle=270]{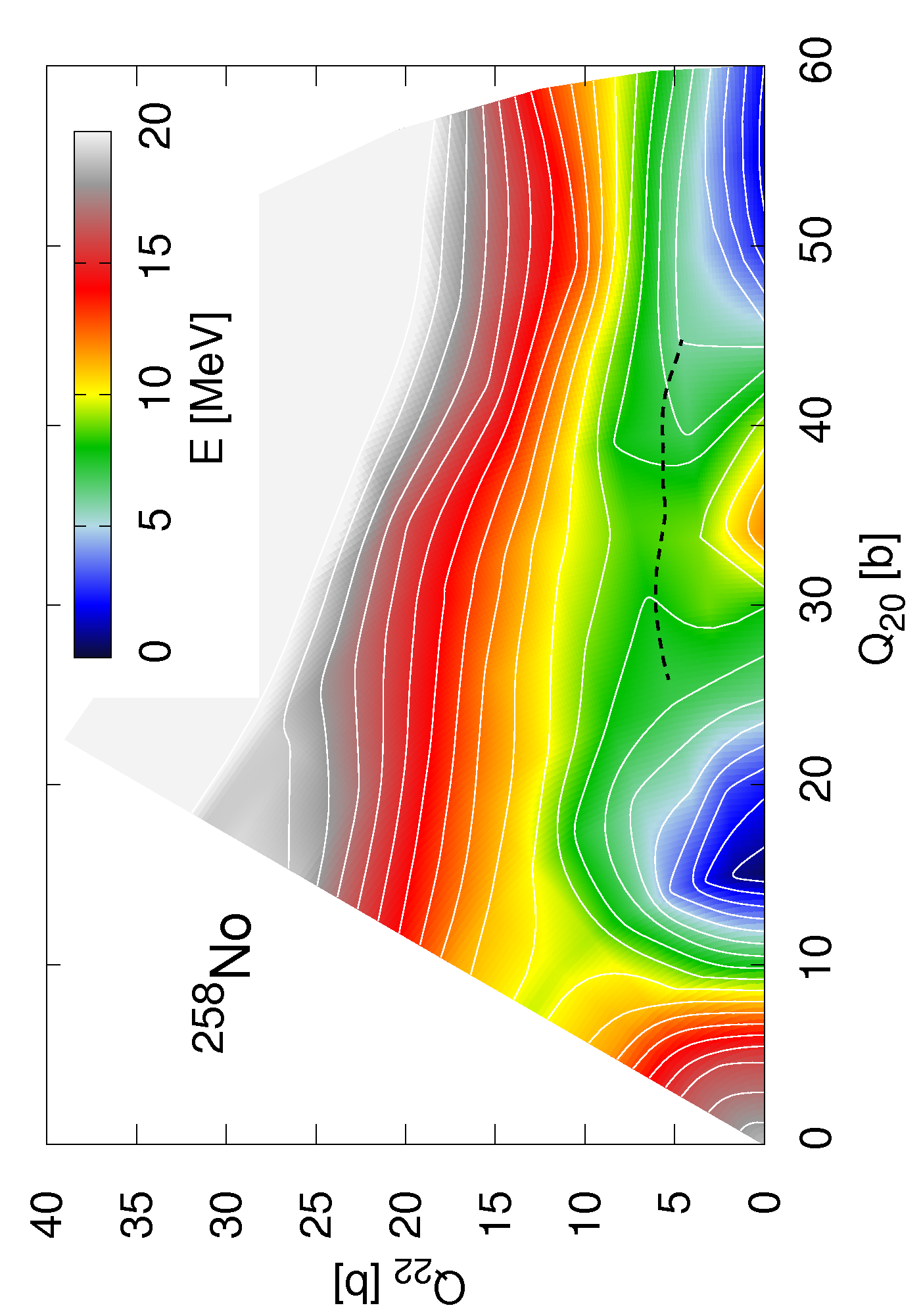}
\caption{ The PES in $Q_{20} -Q_{22}$ plane for $^{252}$Cf (top) and 
$^{258}$No (bottom) are shown as contour and color plots. 
The iso-energy contour lines are plotted every 1 MeV. The 
dashed line shows the fragment of the fission path around the first 
barrier with non-zero triaxial deformation. \label{PEStri}}
\end{figure}

\begin{figure*}
\includegraphics[width=1.9\columnwidth, angle=0]{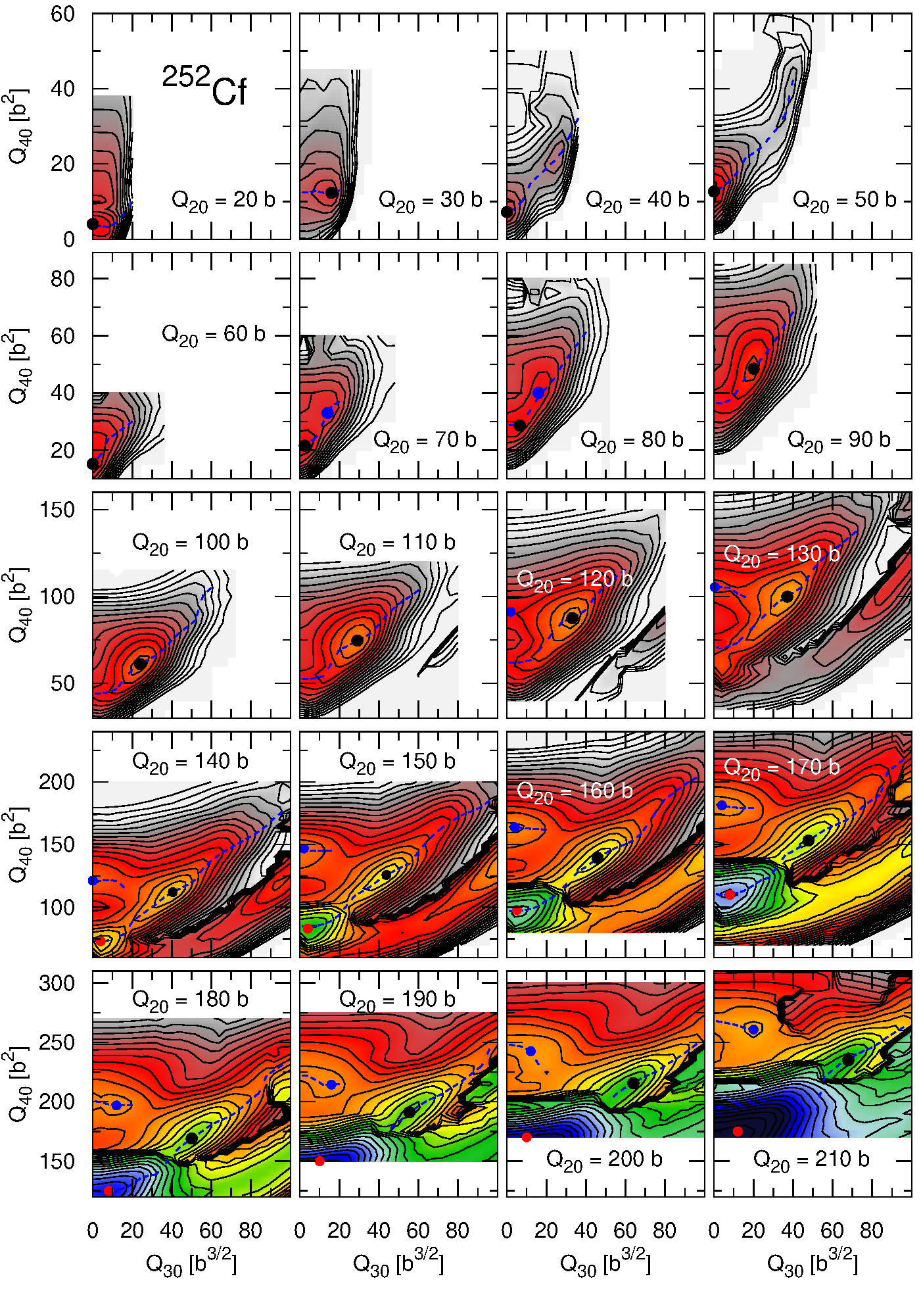}\\
\caption{The PES of $^{252}$Cf in a $Q_{30} - Q_{40}$ plane is shown as 
a contour and color plot. Constant 
energy contours are plotted every 2 MeV. The energy color scale is the same 
as in Fig.~\ref{PEScfno}. See text for details.\label{q4_cf}}
\end{figure*}

\begin{figure*}
\includegraphics[width=1.9\columnwidth, angle=0]{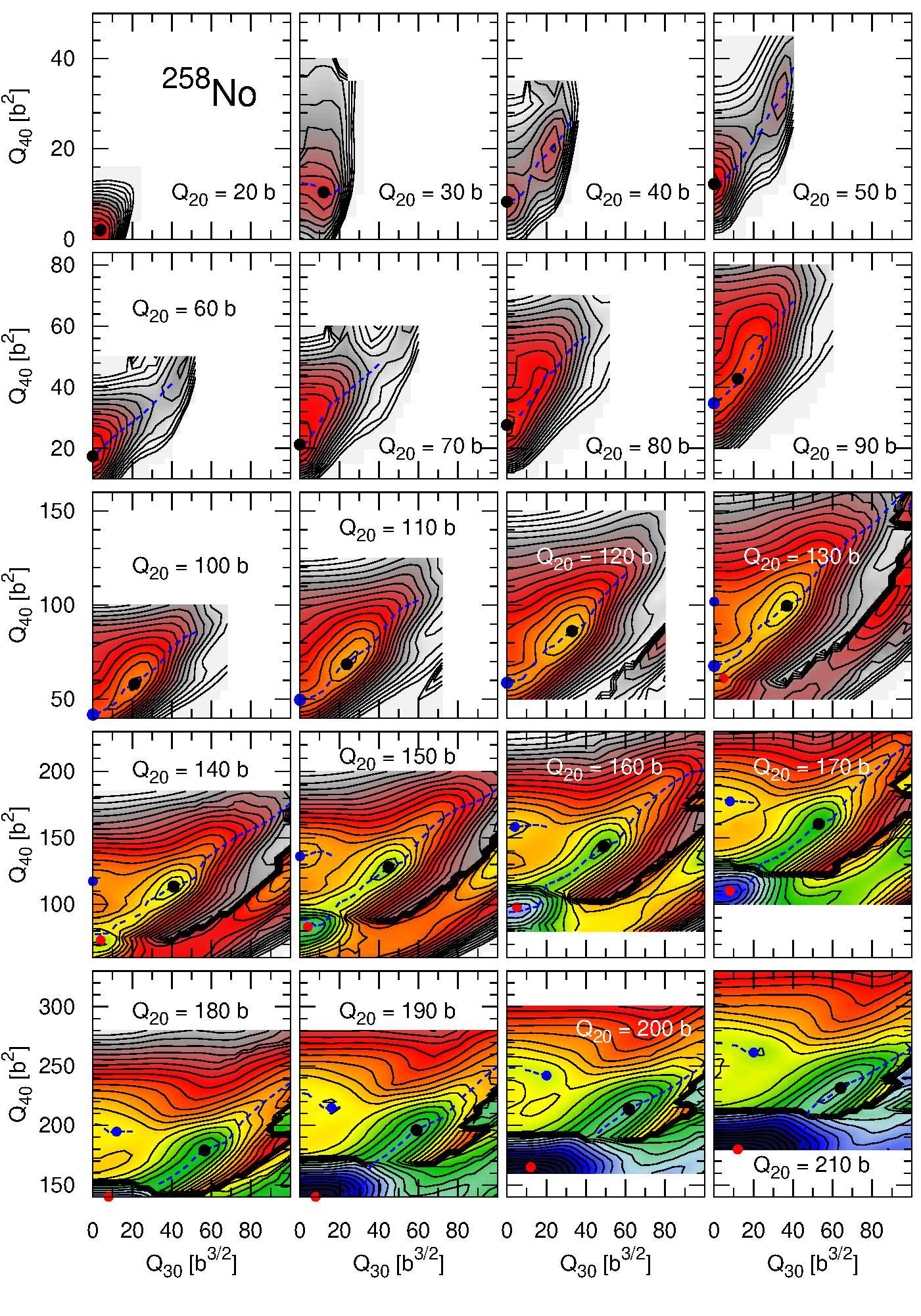}
\caption{The same as in Fig.~\ref{q4_cf} but for $^{258}$No. \label{q4_no}}
\end{figure*}

The theoretical description of fission is based on the analysis of the 
topography of the PES when represented as a function 
of the deformation parameters related to the elongation of the nucleus. More 
elaborated studies also include parameters related to reflection asymmetric shapes 
as they are required for the
description of the asymmetry in fission fragment mass distribution. 
The inclusion of several shape parameters is also helpful
in better characterizing not only the height but also the shape of the fission barrier. 
The shape of the fission PES of $^{252}$Cf and $^{258}$No are 
shown in Fig. \ref{PATHcfno} as a function of the quadrupole 
deformation. In Fig.~\ref{PEScfno}, the PES maps in the 
quadrupole-octupole plane are displayed for the same isotopes. The mesh 
points used to obtain the maps are calculated every $\Delta Q_{20}$= 5 b and 
$\Delta Q_{30}$=4 b$^{3/2}$.

The results presented in Fig.~\ref{PEScfno} have been calculated in the 
axial regime. However, non-axial shapes are crucial to describe the 
region of the first fission barrier correctly, and therefore  we 
present the influence of triaxial deformations on the PES in 
Fig.~\ref{PEStri}. To simplify the calculation, reflection symmetry is 
preserved in these triaxial calculations, and therefore all the shapes 
have zero octupole moment. Usually, it is assumed that the information 
obtained from the aforementioned one- and two-dimensional plots 
together with the collective inertia is sufficient to describe the 
details of the fission process from the ground state till scission. Of 
course, they contain the key elements of the fission process like 
fission barrier heights and fragment mass asymmetry. Nevertheless, 
there are several limitations to this approach. First - the collective 
motion not only explores the minimum energy path (or the minimum action 
path) but also its neighborhood. There is a finite probability that the 
collective wave packet representing the evolution of the system towards  
fission explores regions close to the classical trajectory 
\cite{gutte2005,zdebPhysRevC.95.054608,bulgac2019}. In this respect, 
the width of the valley, or - in other words - the stiffness of the 
potential energy, affects the spread of the collective wave packet and 
therefore the fragment mass distribution. Departure from the classical 
least energy fission path implies considering a broader set of nuclear 
shapes like those obtained by modifying the neck width and length or 
the shape of the pre-fragments. These variations can lead to a change 
of the obtained mas asymmetry at the scission point. Special care 
should be placed on a possible reduction of the neck thickness as it 
determines the scission configuration. 

Another problem is that points on the $Q_{20} - Q_{30}$ map are defined 
as local minima of the energy with given constraints on quadrupole and 
octupole moments. There is no guarantee that the minimum is unique. 
In fact, it was shown that sometimes there might exist multiple local 
minima in the energy \cite{Tsekhanovich2019}. These minima create
different surfaces in the same place of the $Q_{20} - Q_{30}$ map
In such a case, one has to decide which of them should be taken under 
consideration for fission dynamics and, unfortunately, the answer is 
not straightforward. 

Thus, we observe that other degrees of freedom may affect some 
aspects of the theoretical description of fission. It is even the case when a 
two-dimensional map in the self-consistent calculations is created from 
nuclear shapes optimized in the procedure of energy minimization, not 
just created by a two-parameter formula. To better describe and 
understand fission, one has to increase the size of the 
considered space of constraints on which the PES is spanned and look at 
it from a broader perspective. In this way, one should be able to compare 
all the available valleys and deduce which path should be preferred by 
the evolving system.

In order to extend the space of deformation, the most natural and most 
used coordinate is the next term in the multipole expansion: the 
hexadecapole moment $Q_{40}$, responsible for necking 
\cite{Tsekhanovich2019}. By decreasing the value of the hexadecapole 
moment, shapes with a thinner neck are obtained. The alternative option 
is applying the constraint on the neck parameter  $Q_N$ 
\cite{War02,wardastaszczak}. It produces the same effect as the 
hexadecapole moment, but it is more sensitive at the scission region 
and less precise elsewhere.


To visualize a three-dimensional PES is a difficult task that can 
somehow be facilitated if one of the variables is kept fixed and the 
PES for the remaining two variables is plotted as a contour plot. Such 
a procedure has been followed in Fig. \ref{q4_cf} for $^{252}$Cf and 
Fig \ref{q4_no} for $^{258}$No where we show sections of the PES for 
fixed values of $Q_{20}$ as maps in the $Q_{30} -Q_{40}$ space. The 
black, blue, and red dots correspond to the least-energy fission paths 
found on the PES spanned on the $Q_{20} -Q_{30}$ space. The black dots 
correspond to the lowest-energy minimum, the blue ones stand for the 
next-in-energy local minimum, and the red dots indicate the 
post-scission minimum corresponding to two fragment solution. The blue, 
dashed lines show the results of 2-dimensional calculations, where the 
hexadecapole moment is self-consistently given by the minimization of 
the total HFB energy with  double constraints on $Q_{20}$  and 
$Q_{30}$. The thick black line represents scission configurations and 
will be discussed below.

The analysis of the fission barrier in the next subsections will be 
based on all the above mentioned plots of the PES.

\subsection{Detailed description of the PES}

The fission barriers and the PESs obtained for the two nuclei considered 
agree with the expectations for nuclei in the heavy-actinides
region. Both isotopes are prolate in their ground 
states with $Q_{20}= 16$ b. 
On the quadrupole-octupole map of the PES, one can see a fission valley 
heading towards large octupole deformation starting at the ground state. It 
describes super-asymmetric fission strongly related to cluster 
radioactivity \cite{War11, Warda18}.  The minimum of the energy 
corresponding to this valley can be seen in Figs. \ref{q4_cf} and 
\ref{q4_no} at $Q_{20}= 30$, 40 and 50 b and large octupole moments. 
The saddle point in the super-asymmetric valley reaches over 20 MeV and in 
heavy actinides, leads to an exotic decay mode \cite{Warda18}, not observed 
experimentally. We will not discuss this type of fission here.

The first reflection-symmetric barrier is located at around  
$Q_{20}=30 -35$ b. It is well known that triaxial deformation reduces 
its height \cite{DELAROCHE2006103,Afanasjev2010,Chai_2018}. In Figs. 
\ref{PATHcfno} and \ref{PEStri} we can see that the barrier width stays 
unchanged by including triaxiality but the sharp-peaked summit of the axial barrier is cut 
off by 4.7 MeV in  $^{252}$Cf and 3.7 MeV in $^{258}$No. The fission 
barriers including triaxiallity are  7.0 MeV and 7.3 MeV high, respectively. 
The experimental value for  $^{252}$Cf  is 5.3 MeV \cite{britt,Smirenkin}. 
Triaxial deformation of the fission path is relatively small $\gamma \le  
12^{\circ}$. 
At around $Q_{20}= 50$ b the 
nucleus goes back to fission through axial shapes, and triaxiality is 
negligible for larger elongations.

The modification of the landscape by including triaxial deformations is 
non-negligible, but its influence on the spontaneous fission half-life depends
on the value of the collective inertia along the triaxial path as compared to
the axial one. There are indications that triaxiality  
should not affect fission dynamics since the 
energy decrement at the saddle is compensated in the calculation of 
the collective action by the increase of the 
collective inertia and therefore, the tunneling probability is larger along
the axial path~\cite{zhao,DELAROCHE2006103}.  Please note that despite the similarity in barrier 
heights, the experimental fission half-lives of the two isotopes differ by 12 orders of 
magnitude~\cite{Wang1674-1137-36-12-003}. The key to understanding this result,
implying quite different barrier penetrabilities is the existence of a second 
barrier in $^{252}$Cf that disappears
in the  $^{258}$No case, reducing the effective width of the total barrier dramatically.
In consequence, theoretical fission half-lives are $\log_{10}(t_{1/2}/\mathrm{s})=8.74$ 
for $^{252}$Cf 
$\log_{10}(t_{1/2}/\mathrm{s})=-1.94$ for $^{258}$No (calculated in the axial regime). 
These values are less than one order of magnitude away from experimental data.








The second minimum can be found at an energy as low as 1.3 MeV 
above the ground state at $Q_{20}= 50$ b in 
$^{252}$Cf and  0.2 MeV below the ground state at $Q_{20}= 55$ b in 
$^{258}$No. In both isotopes, the shapes of the nucleus are axially 
and reflection symmetric in the second minimum. A well pronounced 
fission valley with non-zero octupole deformation opens up at larger 
elongation in both isotopes. The topography of the PES beyond this 
point is crucial for determining the fission fragments mass 
asymmetry and fission half-lives. The key factor is whether the nucleus would 
prefer to stay in the symmetric fission path or rather turn into the 
octupole valley. The substantial discrepancy between $^{252}$Cf and  
$^{258}$No can be found in the shape of the symmetric fission barrier 
and the shape of the energy surface around $Q_{20}=100 $ b for small octupole 
deformation. The difference is relatively small in absolute values but 
provides important consequences for the fission properties. 

One can notice in the $^{252}$Cf PES that around deformation $(Q_{20}, Q_{30})$ = (100 b, 0 
b$^{3/2}$) a 4.5 MeV high second barrier arises, which blocks the 
symmetric fission channel in this isotope. In fact, from $Q_{20}= 70$ 
b, there is no local symmetric minimum on the PES (there is a peak not saddle in the 
two-dimensional plot), and hence the barrier 
is plotted with a dash-dotted line in Fig \ref{PATHcfno}. In this region, remaining at zero 
octupole moment is energetically unfavorable as the potential energy 
grows 3.2 MeV above the second minimum and the asymmetric valley is 
easily reachable with small energy costs. We can see in Figs. 
\ref{PATHcfno} and  \ref{q4_cf} that the asymmetric valley starts 
already at  $Q_{20}= 70$ b and the second saddle is at 3.1 MeV above the 
ground state (1.8 MeV above second minimum). Moreover, the energy in the 
asymmetric fission path rapidly decreases, reducing the barrier width.

In contrast, the second barrier in $^{258}$No, located at 
$Q_{20}=65$ b, is flat with a height only 0.7 MeV 
above the second minimum (0.3 MeV above the ground state). It is almost 
completely hidden below the ground state and therefore does not contribute in
a substantial way to the half-life, as it was mentioned above.  At 
$Q_{20}= 90$ b the asymmetric valley opens up  without any 
additional barrier (see Fig. \ref{q4_no}). The fission process may 
proceed in the asymmetric mode, but the nucleus may as well stay at 
the reflection symmetric path without energetic costs. This fact explains 
the experimentally observed bimodal fission mass distribution with a small (5\%  abundance) 
component with high kinetic energy and mass symmetric distribution \cite{huletPRL,hulet}.

\begin{figure}
\includegraphics[width=0.99\columnwidth, angle=0]{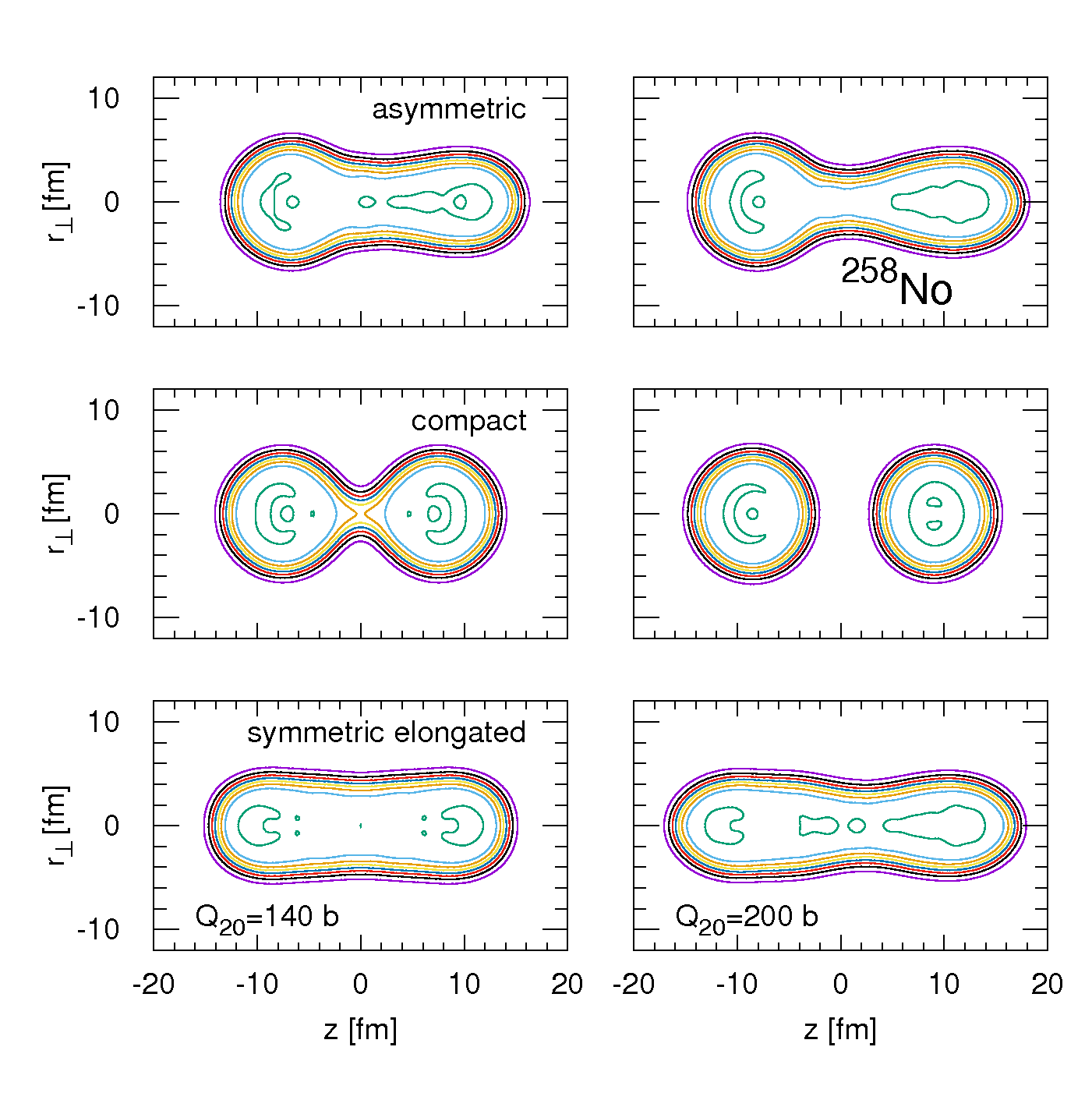}

\caption{The density distribution of $^{258}$No at  $Q_{20}= 140$ b 
(left) and  $Q_{20}= 200$ b (right) in various fission paths:  
asymmetric (top), compact before and after scission (middle) and 
elongated symmetric (bottom). \label{shapes}}
\end{figure}

The topography of the  $Q_{30}$ - $Q_{40}$ planes given for fixed 
quadrupole moment and shown in Figs. \ref{q4_cf} and \ref{q4_no} has 
quite a simple structure up to the region of the second barrier. We can 
see only one -- or at most two -- local minima in parabolic-shaped 
valleys.  In the final phase of the fission process, the PES is far 
much complex. Already at $Q_{20}= 120$ b in both isotopes, we can see 
several branches of the fission valley with
two or three local minima for the same quadrupole moment. The first one is 
octupole deformed. Its shape is depicted in the top panel of Fig. 
\ref{shapes}. It leads to asymmetric fission-fragment mass 
distribution, and therefore we refer to it as an {\em asymmetric} mode. 
The second one with $Q_{30}= 0$ b$^{3/2}$ and small values of $Q_{40} $ 
(around 50 - 70  b$^2$)  is a natural continuation of the symmetric 
fission path. The nuclear shape consists here of two almost spherical 
pre-fragments, see the middle panel of Fig. \ref{shapes}. This one is 
often called a {\em compact} mode of fission.  By stretching this 
configuration, we increase the distance between pre-fragments and make 
the neck thinner up to its disappearance. The compact path is continued 
after scission with two separated fragments. For a two-fragment 
solution, further increasing of the quadrupole moment leads to an 
increase of the distance between daughter nuclei instead of a shape 
change. The minimum of the energy is obtained for a configuration with 
a small mass asymmetry between fragments and, in consequence, a small 
octupole deformation.

The third minimum also corresponds to reflection symmetric shapes, 
but the hexadecapole moment takes much larger values, over 90  b$^2$.  
In this configuration, pre-fragments are not well disentangled. The 
shape is almost cylindrical, {\it cucumber-like} with a small reduction 
of thickness in the region of the neck (bottom panel of Fig. 
\ref{shapes}). This solution is called {\em symmetric elongated} mode 
\cite{staszczak09,Staszczak13}. Small reflection asymmetry can also be
found here, especially for large quadrupole moments. The corresponding 
fission path is the highest in energy, and it survives up to a very large 
elongation of the nucleus as the neck is not formed and a rupture of the 
system is not possible without a substantial energy cost.  Calculations 
with the Skyrme energy density functional
suggest that this fission path is energetically comparable with an 
asymmetric path around the second barrier in the nuclei from the region 
of heavy actinides and may play an important role in fission 
\cite{staszczak09,Staszczak13}.

We should stress here that the crossing of fission paths in Fig. 
\ref{PATHcfno} cannot be interpreted as a possible place of bifurcation 
or configuration change. This is a typical example of a fake saddle 
point \cite{DUBRAY2012}. The fact that two or even three lines have 
the same energy for the same quadrupole moment does not mean that they 
represent the same or similar shapes of the nucleus. As one can see in 
Fig. \ref{shapes} the differences in the nuclear density distribution 
between fission paths are usually huge.

The existence of two local minima and surrounding local valleys (e.g. 
compact and symmetric elongated) at the same quadrupole and octupole 
deformation leads straight to the multiplication of the surfaces on the 
traditional elongation--mass asymmetry maps, like in  Fig. 
\ref{PEScfno}. The blue dashed lines in Figs. \ref{q4_cf} and 
\ref{q4_no} indicate solutions corresponding to the local minimum of 
the energy for fixed  $Q_{20}$ and $Q_{30}$, i.e. these data could be 
used to create the PES map in the ($Q_{20}$, $Q_{30}$) space. 
Mixing values coming
from different valleys may lead to an ambiguous and often erroneous interpretation of the 
calculated results which might eventually suggest contradictory conclusions.
As the surfaces are often similar in energy, the choice of the 
surface would likely be a matter of a random selection by a numerical 
procedure or an arbitrary decision of the researcher. There is also the 
risk of an accidental change of the surface. This effect could  
easily be omitted in the analysis of fission as the graphical plotting programs are 
likely to smooth out sharp ridges. The application of the density 
distance parameter described below in Subsection \ref{dd} can be useful to 
prevent these non-physical interpretations. 

In the analysis of Figs. \ref{q4_cf} and \ref{q4_no} various scenarios 
for reaching pre-scission configuration can be envisaged. The scission 
line is visible in these plots for $Q_{20}\ge 120$ b as a black line 
(consequence of plotting many closed energy contour lines) separating 
single shapes (above) from two-fragments configuration (below). The 
rapid change on the energy is a consequence of the strong dependence of 
the neck with multipole moment parameters: slight changes lead to a 
strong reduction of the neck and eventually to the splitting of the 
nucleus (see discussion below).

Beyond the region of the second fission barrier ($Q_{20}=100$ b), the 
PES goes down towards scission. The pre-scission line, i.e. the line of 
the most elongated shapes before the rupture of the neck, is clearly 
seen at Fig. \ref{PEScfno} as a few MeV high cliff that separates the 
fission valley from the two-fragments configuration. There are several 
ways of reaching a scission configuration: The first is through a 
symmetric compact mode. Increasing elongation leads to a relatively 
smooth shape evolution from pre- to post-scission configuration. 
Nevertheless, even if a energy fall is not noticed, an abrupt change in 
the nuclear density distribution around the neck region is visible 
(e.g. see \cite{Warda_2015} for the  $^{258}$Fm case). The two 
dimensional $Q_{30}-Q_{40}$ sections of the PES between $Q_{20}=120$ b 
and 130 b in $^{252}$Cf and between 150 b to 160 b in $^{258}$No do not 
differ substantially among them despite a few MeV energy drop in the 
minimum of the valley. It indicates that the pre- and post-scission 
density distributions of the whole system are relatively similar 
despite the neck rupture. Another, most intuitive, way of reaching a 
scission point from the asymmetric fission valley is by going along the 
fission path (marked with the blue dashed line in Figs. 
\ref{PATHcfno} and \ref{PEScfno} and
black dots in Figs. \ref{q4_cf} and \ref{q4_no})  to the largest 
possible quadrupole moment. As the energy decreases simultaneously with 
increasing elongation, the largest gradient indicates this is the most 
probable fission scenario. The large mass asymmetry $A_H/A_L \sim 
140-142/112-116$ obtained before the neck rupture in this point 
corresponds to the experimentally observed most probable fission 
fragment asymmetry.  

The two scenarios presented above for the evolution of the nuclear 
shape assume that the system remains on the fission paths presented in 
Fig. \ref{PATHcfno}. Therefore, only one or two particular shapes of 
the nucleus are taken at scission. Such analysis can explain the 
fission modes observed in the experiment, but it is insufficient to 
reproduce the details of the fission fragment mass distribution. The 
reason is that on its way from the saddle to scission, the collective 
wave packet may explore different configurations away from the lowest 
energy fission path, still fulfilling the condition of the descending 
energy but not with the largest gradient. In this way, every 
configuration of the scission line in the neighborhood of the 
asymmetric fission valleys which are visible in Fig. \ref{PEScfno} is 
accessible, of course with a reduced probability 
\cite{zdebPhysRevC.95.054608}. These configurations can be observed in 
Figs. \ref{q4_cf} and \ref{q4_no}. For $Q_{20}\ge 140$ b the 
post-scission compact minimum in the south-west corner of each panel 
has got lower energy than on the asymmetric path. The latter valley 
seems to be soft in the direction towards the first one, and the ridge 
separating them does not exceed 2 MeV.  Those figures indicate that the 
exit points from the asymmetric fission valley are available already at 
relatively small octupole moments and with a smaller asymmetry of 
nuclear shapes than at the end of the asymmetric fission path.

Two additional aspects of this scenario must be pointed out. First, a 
scission point is accessible at much less elongated nuclei than at the 
end of the asymmetric fission path. In some lighter nuclei, where a 
scission point is above the ground state energy, this would reduce the 
width of the barrier. In consequence, tunneling probability and the 
half-lives can be substantially shortened \cite{Tsekhanovich2019}. 
Second, at smaller quadrupole moment, an exit point from the asymmetric 
valley at low octupole moment may be ended not in fusion valley but in 
the compact fission valley before scission. This is another mechanism 
of feeding the symmetric mode of fission. Thus, the dynamical 
calculations of the fission mass yields for $^{252}$Cf, where only 
quadrupole and octupole deformations were taken under consideration, 
showed a small contribution for the symmetric mass 
division~\cite{zdeb}.

\subsection{Surface discontinuity and density distance \label{dd}}

\begin{figure}
\includegraphics[width=0.7\columnwidth, angle=270]{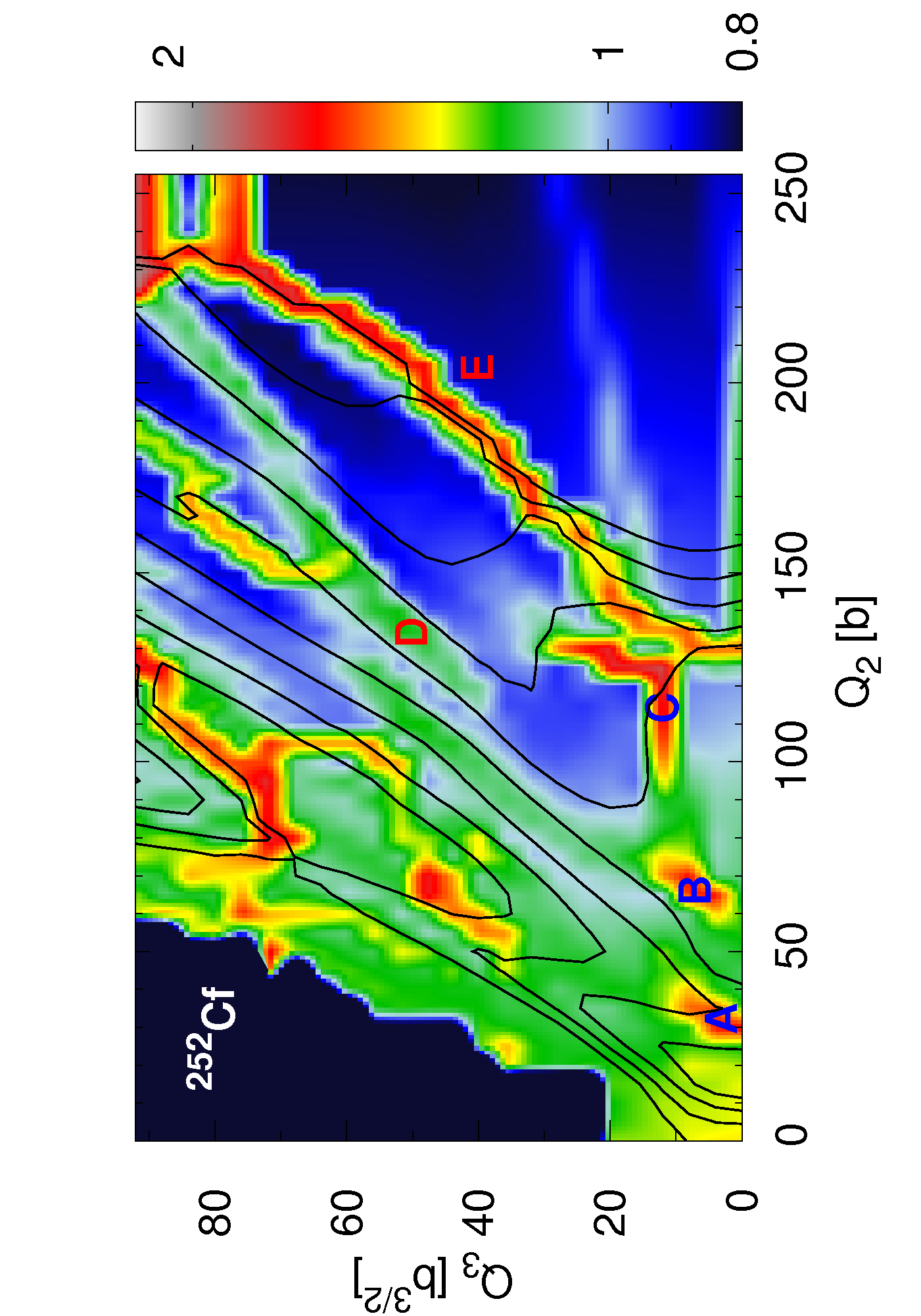}\\
\includegraphics[width=0.7\columnwidth, angle=270]{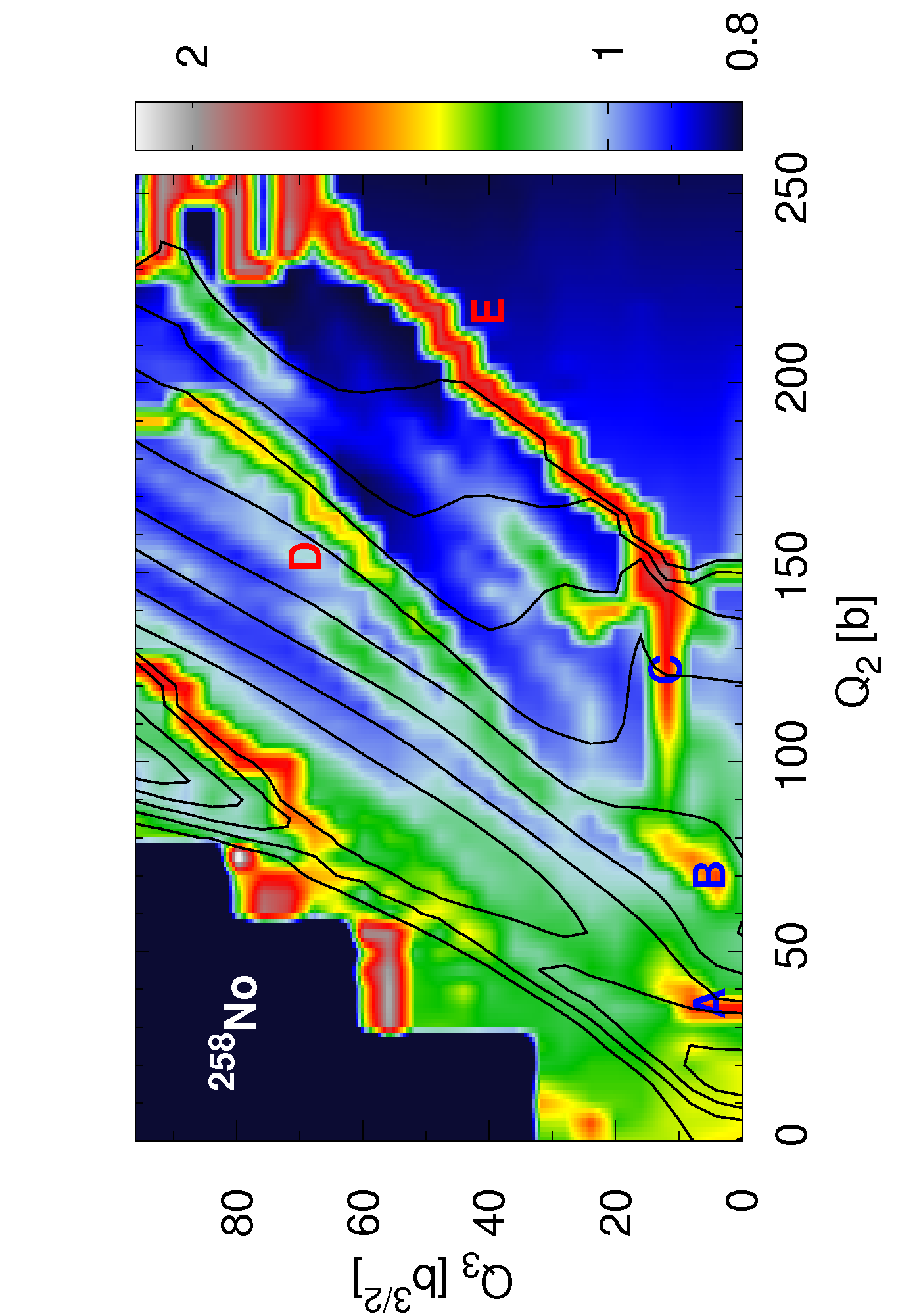}
\caption{The density distance function $D_{\rho\rho'}$ (in logarithmic 
scale) given by Eq. (\ref{dens_dist}) between two neighboring 
configurations in a deformation space $Q_{20}-Q_{30}$ of $^{252}$Cf 
(top) and $^{258}$No (bottom). Contour lines for the energy are plotted every 5 MeV 
for better identification of the different regions of the PES. \label{disc_cfno}}
\end{figure}

\begin{figure}
\includegraphics[angle=270, width=0.9\columnwidth]{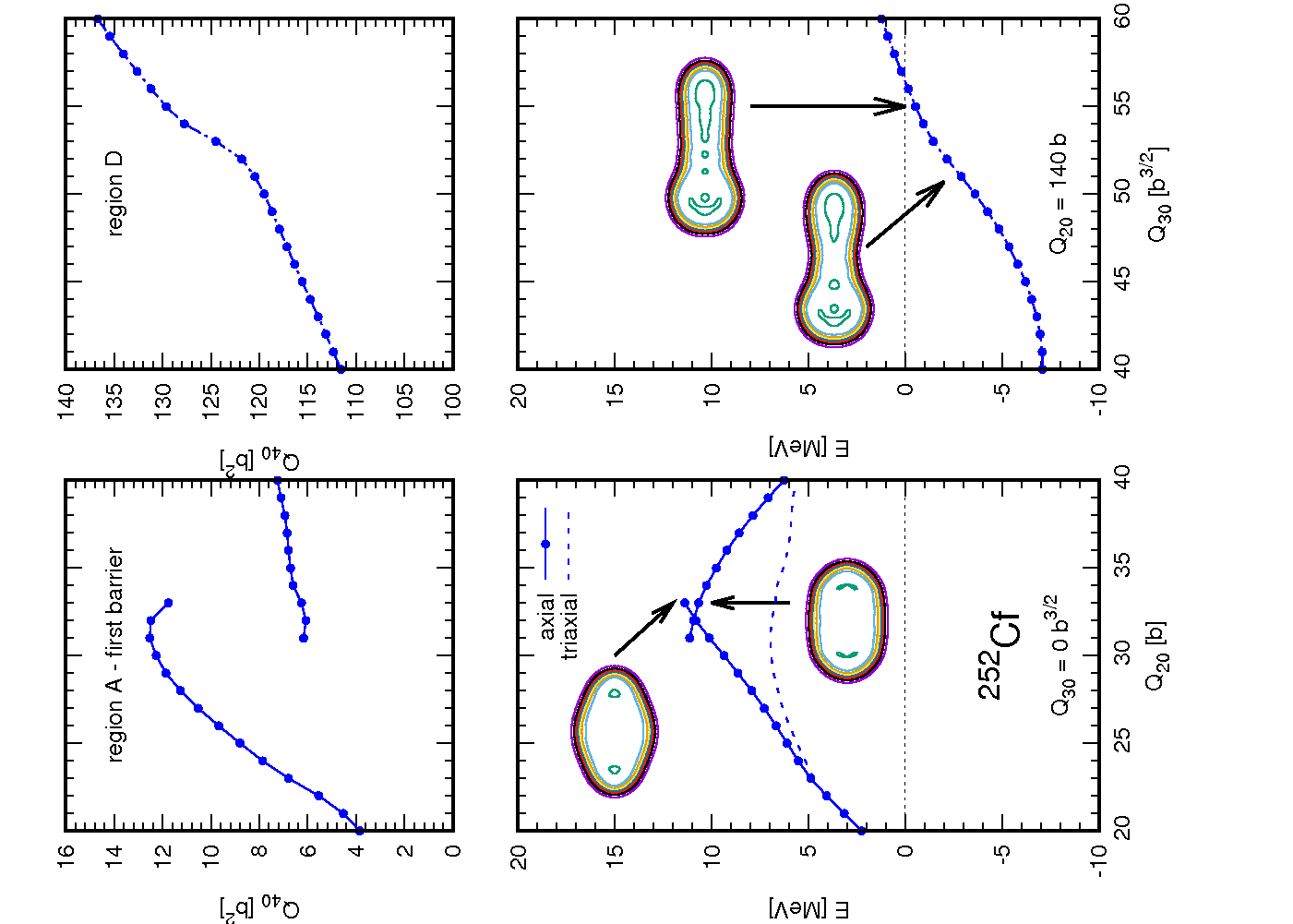}
\caption{Fragments of the first barrier and asymmetric valley at fixed 
$Q_{20}=140$ b of  $^{252}$Cf.\label{sy-asy3}}
\end{figure}

\begin{figure*}
\includegraphics[angle=270, width=1.9\columnwidth]{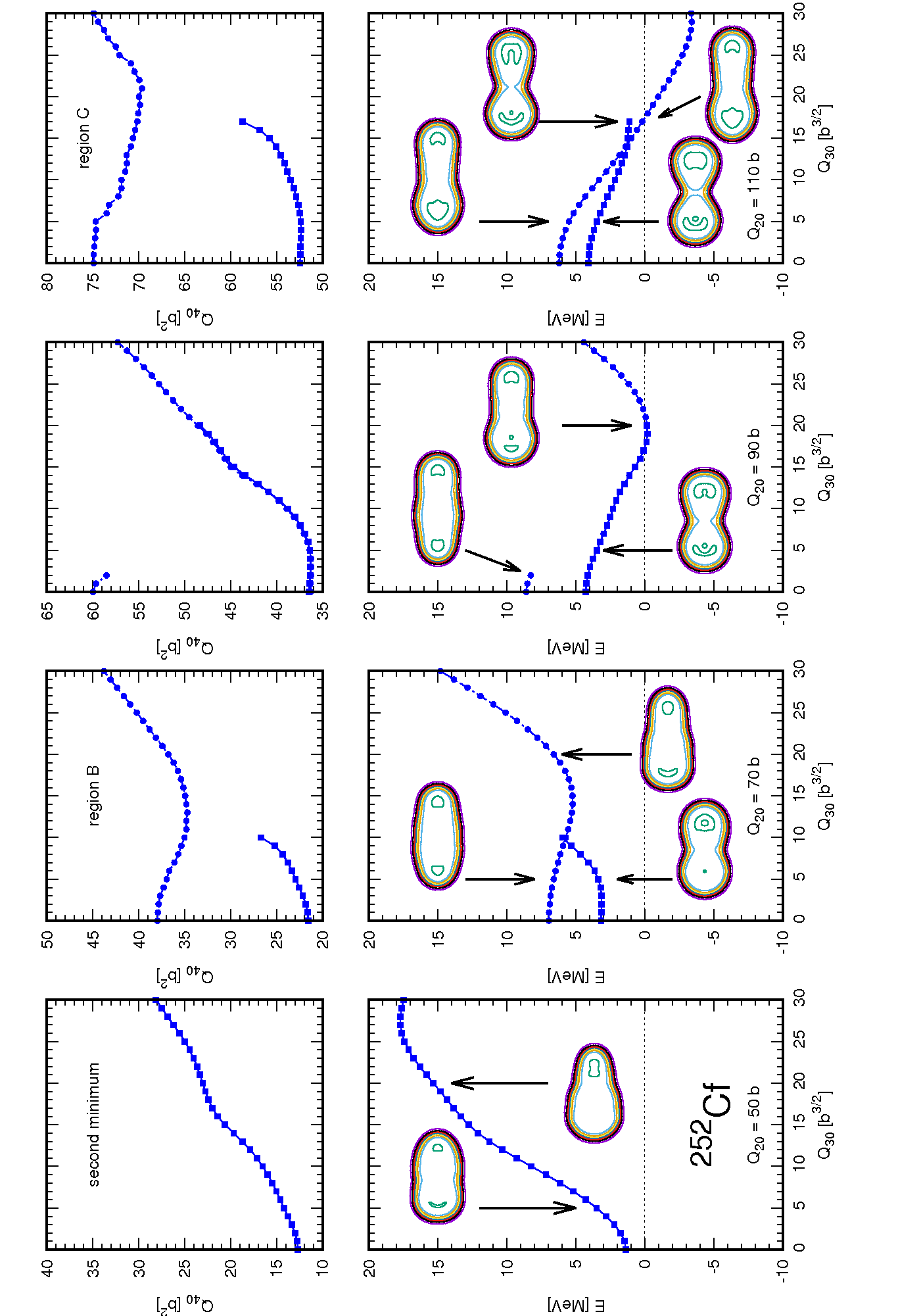}
\caption{Fragments of the PES of  $^{252}$Cf  at fixed $Q_{20}=50$ b, 
70 b, 90 b and 110 b.\label{sy-asy1}}
\end{figure*}

\begin{figure*}
\includegraphics[angle=270, width=1.9\columnwidth]{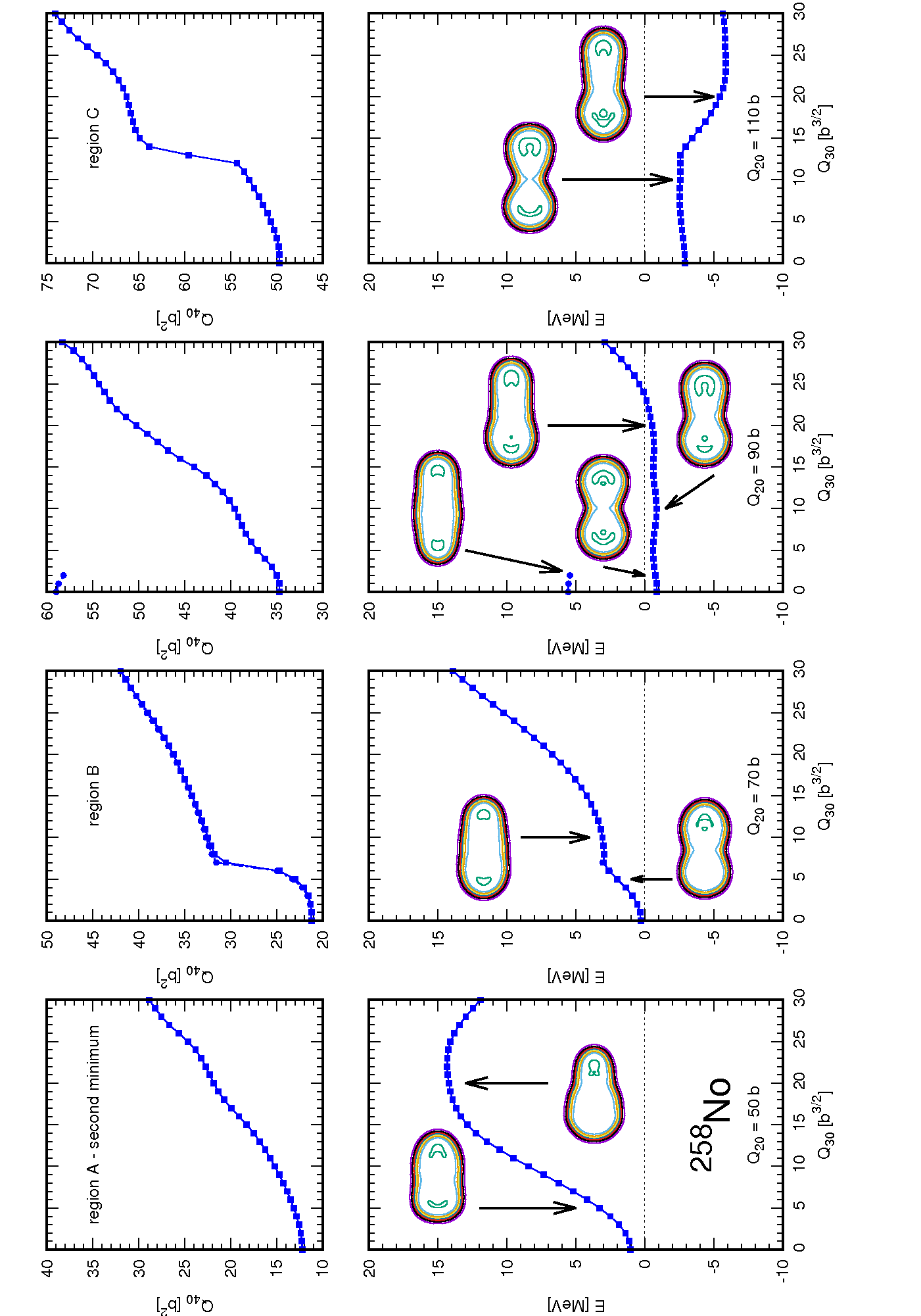}
\caption{The same as  in Fig. \ref{sy-asy1}, but for  $^{258}$No.\label{sy-asy2}}
\end{figure*}

\begin{figure}
\includegraphics[width=0.5\columnwidth, angle=270]{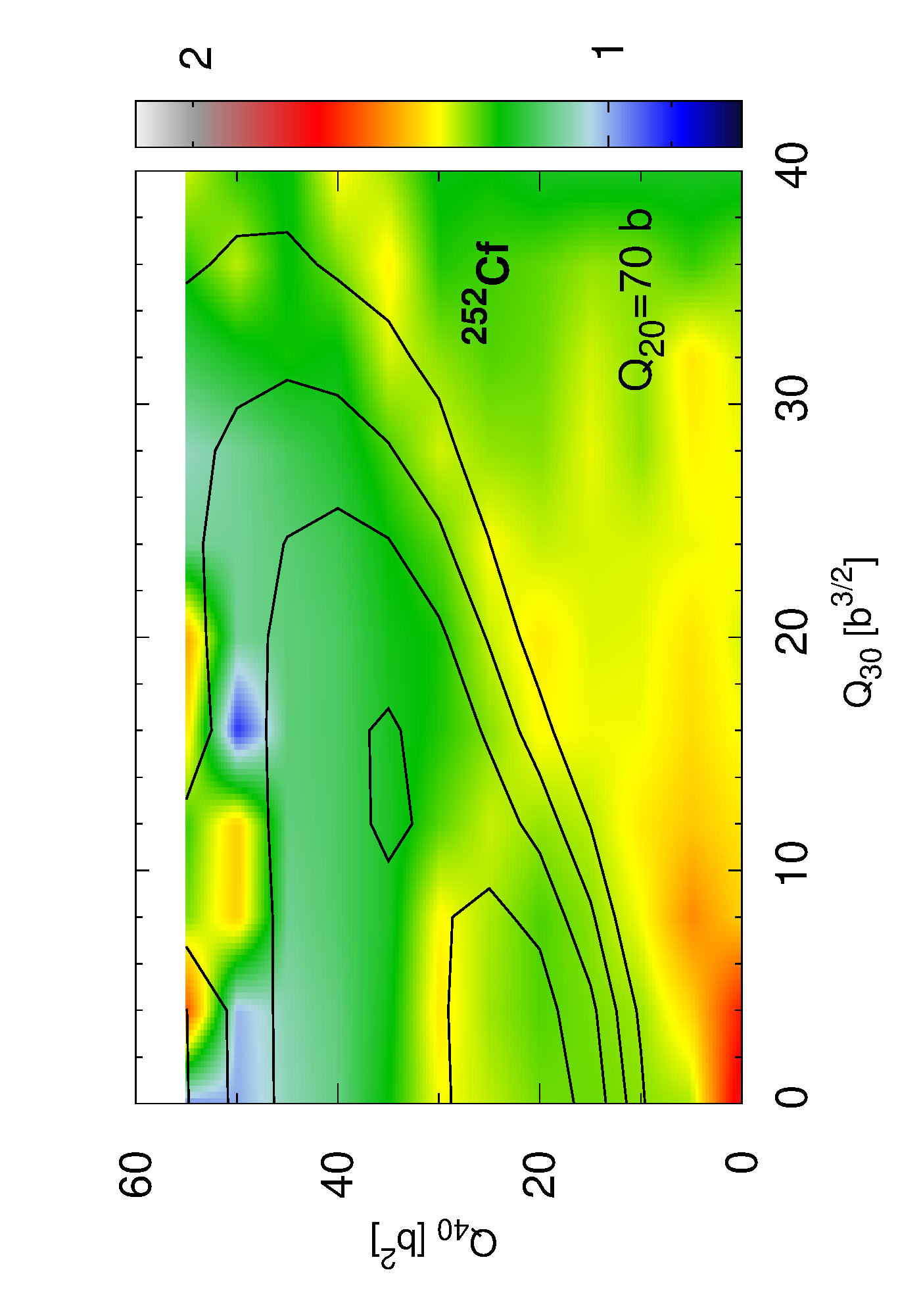}\\
\includegraphics[width=0.5\columnwidth, angle=270]{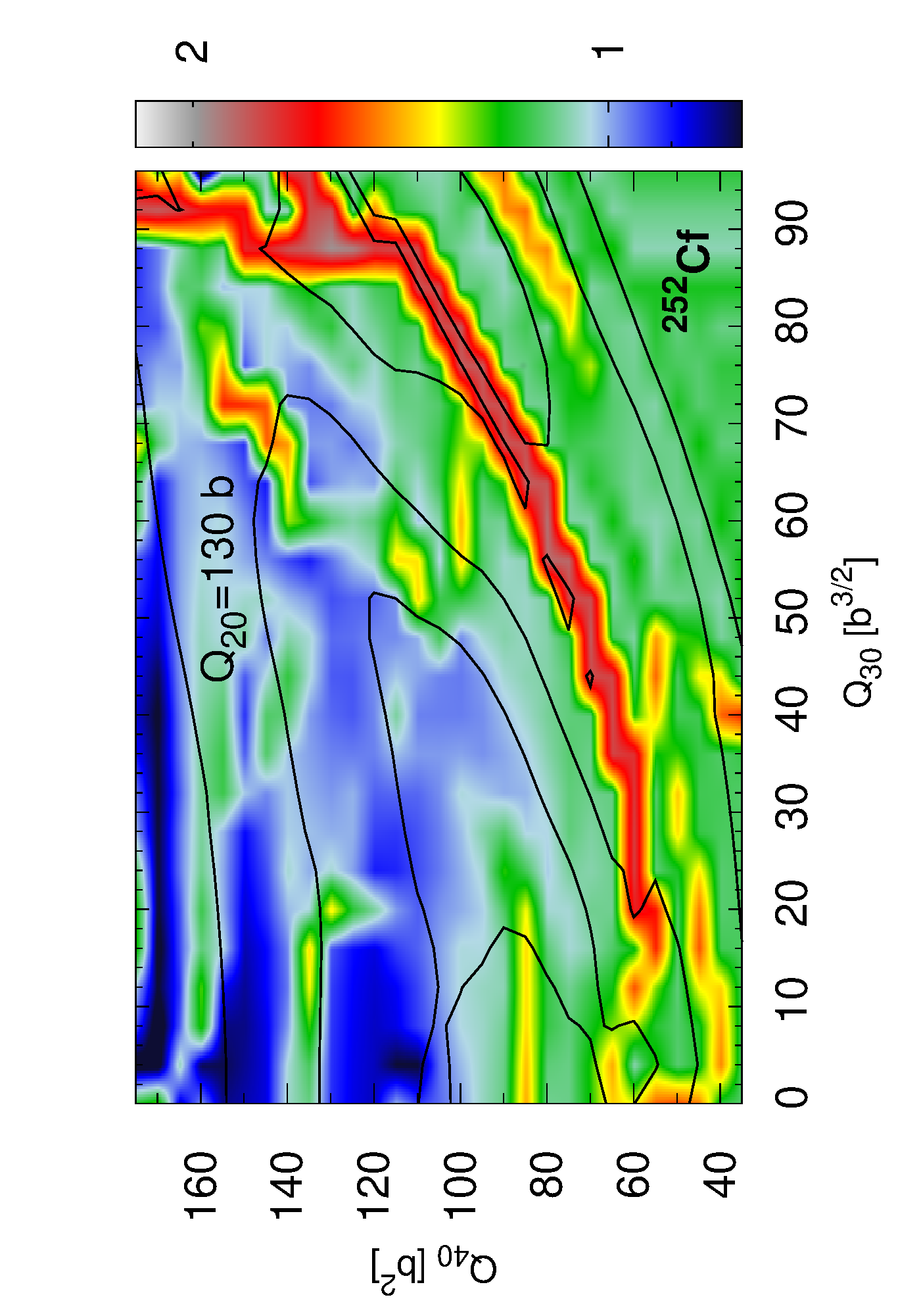}\\
\includegraphics[width=0.5\columnwidth, angle=270]{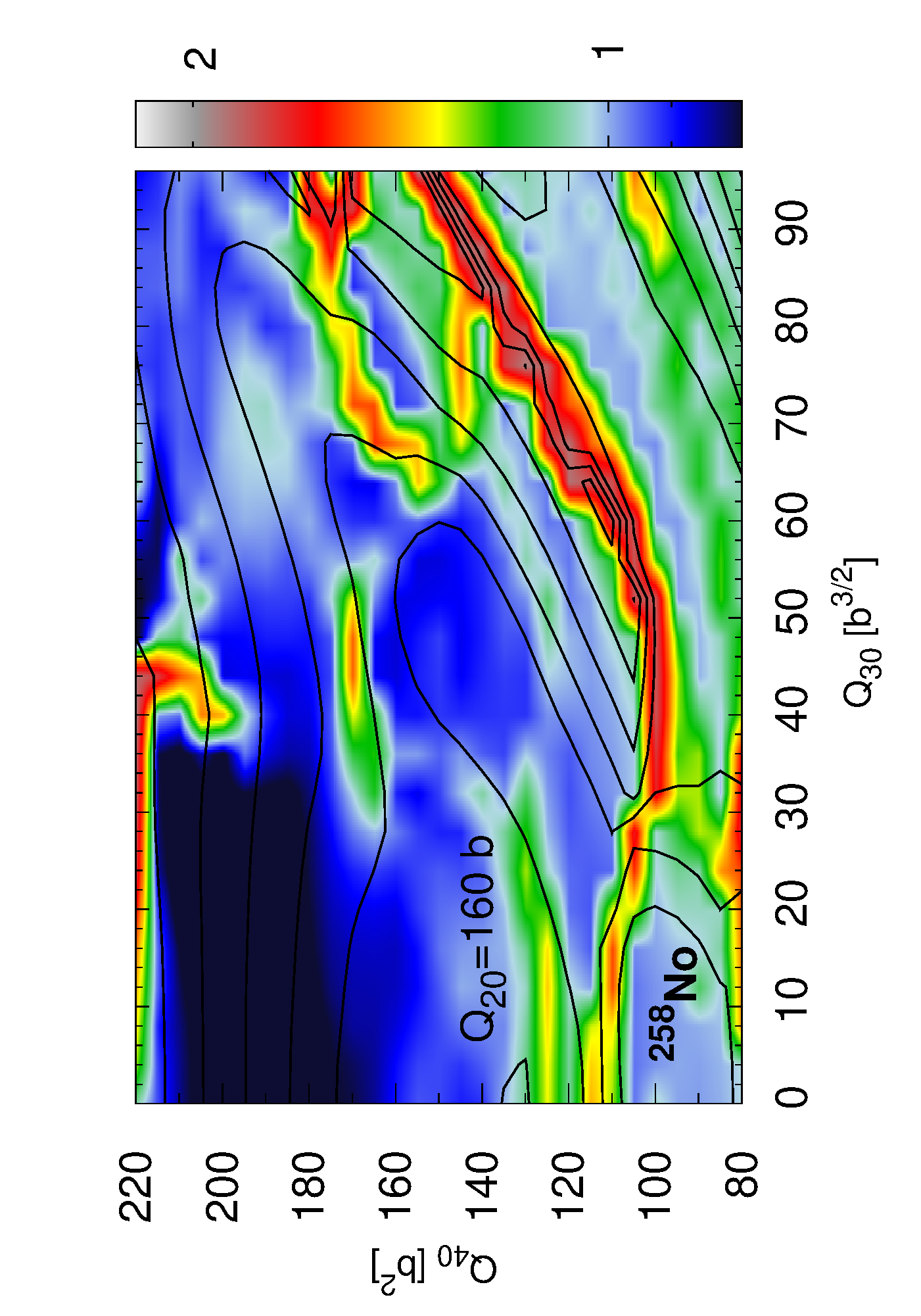}

\caption{The same as in Fig. \ref{disc_cfno}, but for the PES 
at fixed $Q_{20}$ in a deformation space $Q_{30}-Q_{40}$ of 
$^{252}$Cf at $Q_{20}=70$ b (top),  at $Q_{20}=130$ b (middle) and 
$^{258}$No at $Q_{20}=160$ b (bottom).  Contour lines of equal energy are plotted 
every 5 MeV for better identification of the different regions of the PES.  
\label{disc_q3q4}}
\end{figure}

As mentioned in Section \ref{THEORY}, the density distance is an 
appropriate quantity to check the consistency of the PES and pinpoint 
possible discontinuities of the surface that may lead to false 
conclusions concerning fission dynamics. In Fig. \ref{disc_cfno} 
we have plotted, for each point in the $(Q_{20}, Q_{30})$ plane, the largest 
of the two density distances computed for the two configurations 
$(Q_{20}+\Delta Q_{20},Q_{30})$ and $(Q_{20},Q_{30}+\Delta Q_{30})$. 
The figures look pretty similar for both isotopes and a few regions of 
large $D_{\rho\rho'}$ can be identified:

\begin{enumerate}[A -]
\item first barrier;
\item around the second minimum;
\item between symmetric and asymmetric valley at $Q_{30}\sim 15$ b$^{3/2}$;
\item bordering asymmetric valley at large octupole deformation;
\item scission line.
\end{enumerate}
We will denote these regions as ``regions of discontinuities" due
to the large difference in densities corresponding to neighboring points 
indicating on possible abrupt change of configuration.

The first region of discontinuity (A) is encountered already at the 
first barrier. The nucleus has a double cone (diamond) shape on the up-going part 
of the barrier, whereas a two-center structure is created beyond the 
maximum of the barrier, as it can be seen in the left panels of Fig. 
\ref{sy-asy3}. Both density distributions are clearly different with a 
significant jump in hexadecapole moment. A sharp peak of the first 
fission barrier also indicates a sudden change of configuration, and the 
fake or missing saddle mentioned above is usually present 
\citep{DUBRAY2012} because two solutions can be found for the same 
quadrupole moment around the peak. The density distance calculated in 
this region does not take into account the influence of triaxiality 
discussed in Subsection \ref{subsPES} that eliminates discontinuities.

In the right panels of Fig. \ref{sy-asy3}, the region (D) of enhanced 
density distance is observed. As we can see, there is no discontinuity 
here but only a change of the hexadecapole moment can be 
observed in this region. The source of this kink can be explained by 
looking at Figs. \ref{q4_cf} and \ref{q4_no}. The valleys plotted 
in the $Q_{30}-Q_{40}$ planes around $Q_{20}=140$ b 
lie along a more or less straight line with a slope of the order of 1. 
We determine the localization of the 
bottom of these valleys by minimization with a constraint on the 
octupole moment (marked by blue dashed lines). This procedure sometimes 
does not give the correct value that is obtained by following 
the direction of
the gradient along $Q_{30}$ and $Q_{40}$ that determines the bottom of the
valley. As a consequence of the improper determination of the bottom of the
valley using a constraint on the octupole moment, large increases in 
the value of  $Q_{40}$ can be noticed between $Q_{30}= 
50$ b $^{3/2}$ and  55 b $^{3/2}$ in Fig \ref{sy-asy3}. No real 
discontinuity of the surface is found here, but rather a problem with 
the precise numerical evaluation of the bottom of the valley and therefore
no important physics is missed here. 

The regions (C) and (D) should be discussed together as they separate 
the compact symmetric and the asymmetric 
fission valleys. A first look at the PES in  Fig.  \ref{PATHcfno} reveals a seemingly smooth 
surface: no sharp ridges or sudden energy changes.
Therefore, it comes as a surprise to find a discontinuity here. To solve 
this puzzle, one has to use a magnifying glass: in  Figs. \ref{sy-asy1}  
and \ref{sy-asy2} we have plotted a blown-up view of the sections of 
both valleys in $^{252}$Cf and $^{258}$No, respectively. The step size 
in these calculations was reduced to $\Delta Q_{30}= 1$ b $^{3/2}$ with 
starting point from the nearest mesh point with smaller or larger 
octupole moment.

In the left columns of Figs. \ref{sy-asy1}  and \ref{sy-asy2}, the PES 
in the second minimum is plotted. In this case, the energy grows up 
smoothly with 
increasing mass asymmetry. Only a small bending in octupole moment can be 
noticed. The PES picture at $Q_{20}=70$ b, in the region (B) of 
discontinuity, is completely different, see panels in the 
second column.  We find a kink in the energy curve at a relatively small octupole 
deformation. At this point, a clear change of 
configuration takes place as the neck, clearly visible for near 
reflection symmetry shapes, is not pronounced anymore in more asymmetric shapes. Also, 
the hexadecapole moment jumps from $Q_{40}=25$ b$^2$ to 35 b$^2$. It is 
easy to distinguish here the compact valley from the elongated asymmetric one. 
Moreover, in $^{252}$Cf, the asymmetric valley can be extended towards
lower octupole moments up to zero. This is the germ of the symmetric 
elongated fission valley. Discontinuity in the region (B) is not a 
significant problem in the description of fission as a transfer from the 
second minimum to the asymmetric valley requires a few additional MeV of 
energy, which practically blocks such evolution.

Increasing the  quadrupole moment, we find a space between regions (B) and 
(C) in Fig. \ref{disc_cfno}. Its impact is also visible in Figs. 
\ref{sy-asy1}  and \ref{sy-asy2} at $Q_{20}=90$ b. Here, the PES as a function 
of $Q_{30}$ is continuous again. Hexadecapole moment gradually increases 
indicating a smooth connection between the compact and asymmetric 
valleys. The transfer between configurations is possible as a consequence 
of decreasing the energy of the asymmetric minimum. The tiny part of the 
symmetric elongated surface can be noticed only very close to 
reflection symmetric shapes.

The rightmost panels of Figs. \ref{sy-asy1} and \ref{sy-asy2} describe 
the region (C) of discontinuity. Despite the same values of the 
quadrupole and octupole moments and the similarity of energies, the 
hexadecapole moments and the shapes of the nucleus are considerably distinct 
in both configurations. In the compact symmetric mode, two 
pre-fragments are separated by a thin neck. In the asymmetric mode, the 
density distribution is more uniform along the symmetry axis. In both 
isotopes, the compact mode is limited to hexadecapole moments in the 
range from $Q_{40}=50$ b$^2$ to 60 b$^2$ whereas the asymmetric mode is 
described by much higher values, over 65 b$^2$. 
In $^{252}$Cf, the 
compact valley created by increasing the octupole moment ends 
with a sudden drop into the asymmetric valley. 
In this nucleus, by
decreasing octupole moment in the calculations of consecutive mesh 
points in the asymmetric valley, one may reach zero octupole moment in 
a configuration characteristic for the elongated symmetric valley. The 
transition from the compact to the asymmetric valley is an analogue of 
the scission line discontinuity described below. In  $^{258}$No both 
surfaces meet at the same energy, and the elongated symmetric part can not 
be determined at  $Q_{20}=110$ b. In region (C), in both isotopes, transfer to the 
asymmetric valley is energetically favorable.

The transition between configurations in this region is usually 
overlooked. It is easy to link both valleys when the mesh points are 
scattered by 4 or 5 b$^{3/2}$, which is usually a reasonable choice 
while producing the PES maps. The similarity of the energy slopes 
as well as the fact that increasing 
octupole moment on the compact valley beyond its end, leads to a 
solution in the asymmetric one (by applying the self-consistent energy 
minimization) enhances the chances of making mistakes. The discontinuity in the region (C) signals the 
frontier between the compact mode surface in the symmetric 
second minimum region and the symmetric second barrier.

Looking at the energy section maps in Figs. \ref{q4_cf} and 
\ref{q4_no} we find that the discontinuity between symmetric and asymmetric 
surfaces discussed above is a good example of missing saddles as discussed by Dubray and Regnier
\citep{DUBRAY2012}. In three-dimensional PES, the problem of a rapid 
change of configurations disappears.

Finally, a commonly known discontinuity is localized in the region (E) 
of the scission line. There is a huge difference in pre-scission and 
post-scission configurations.  It is very hard or even impossible to 
find a continuous link by using constraints on multipole moments to 
control the nuclear shape. Applying constraints on the neck parameter 
$Q_N$ may help to provide the continuous surface on the scission line 
\cite{War11,wardastaszczak}.

Let us discuss now the source of the scission line {\it cliff} 
obtained in the self-consistent calculations. The asymmetric valley in 
two-dimensional space is build of configurations, which are local 
minima along {\it all} the directions orthogonal 
to the ones of the constraints. 
In consequence, decreasing the neck thickness (or hexadecapole moment)
leads to increasing the energy of the system even though the ruptured 
nucleus is energetically favorable.
The height of the barrier separating the 
asymmetric path from the no-neck solution reduces to zero with 
increasing quadrupole moment. Beyond the pre-scission line, any shape of the 
nucleus is unstable against neck rupture, i.e. the energy 
monotonically decreases with decreasing neck thickness. The  
energy minimization procedure cannot find a stable solution with a neck. 
The gradient of the energy directs the system 
towards post-scission configuration with much lower energy in the 
self-consistent process. Of course, 
the cliff on the scission line does not mean that in Nature the neck 
disappears instantly, but only that a further thinning of the neck 
should occur without increasing the elongation of the system.

The concept of density distance can also be applied to the PES given as 
a function of the octupole and hexadecapole moments for fixed $Q_{20}$ 
values. An example of the results is presented in Fig. \ref{disc_q3q4}. 
Here again, the discontinuity at the scission line is clearly visible. 
The density distance is also enhanced in some other regions of the PES 
quite randomly scattered on the surface. Its values are relatively 
small in comparison to the scission line ones. The only characteristic 
region of larger density distance separates elongated symmetric valley 
from the compact one.

\subsection{Multiple solutions}

\begin{figure}
\includegraphics[angle=270, width=0.9\columnwidth]{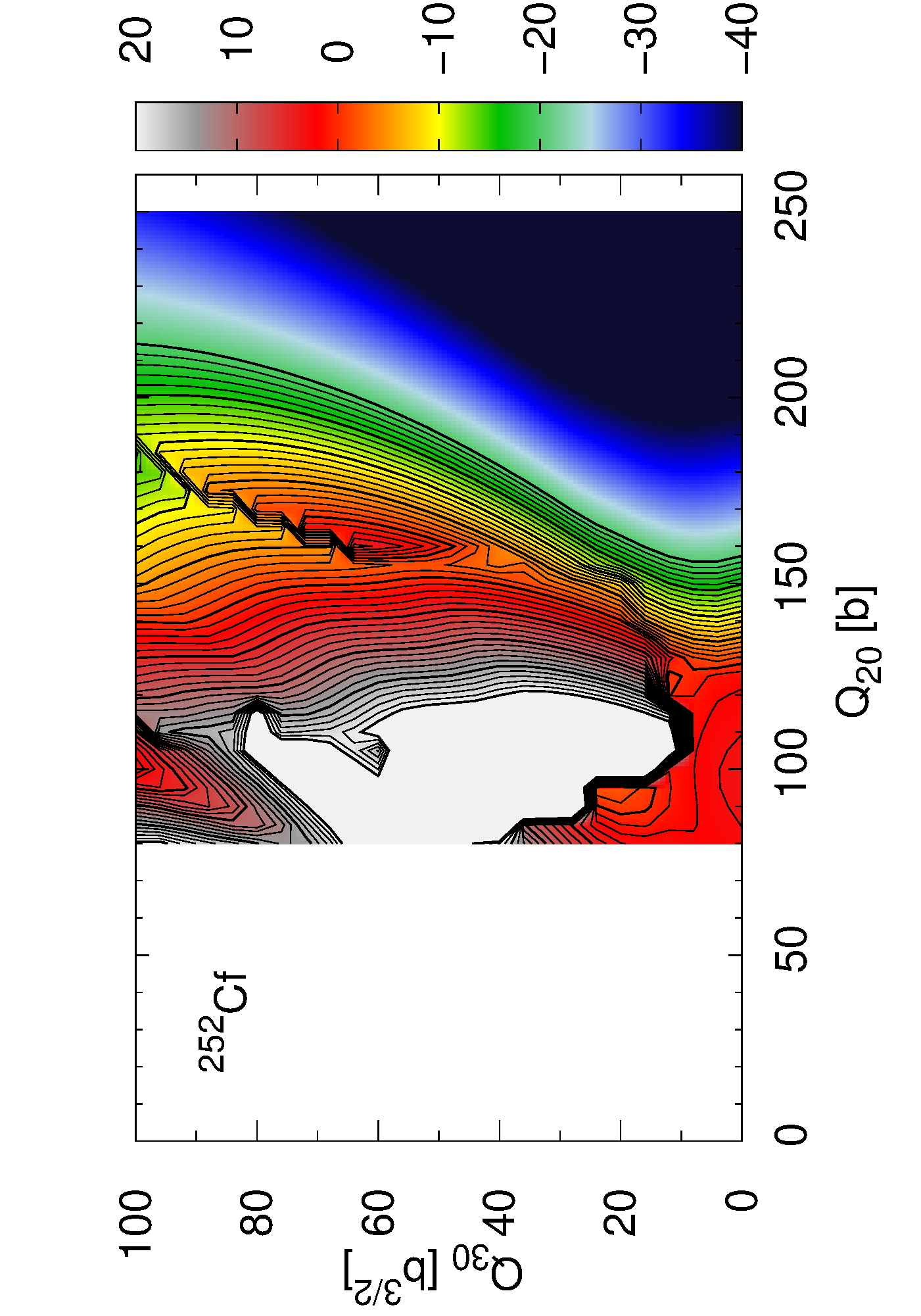}\\
\includegraphics[angle=270, width=0.9\columnwidth]{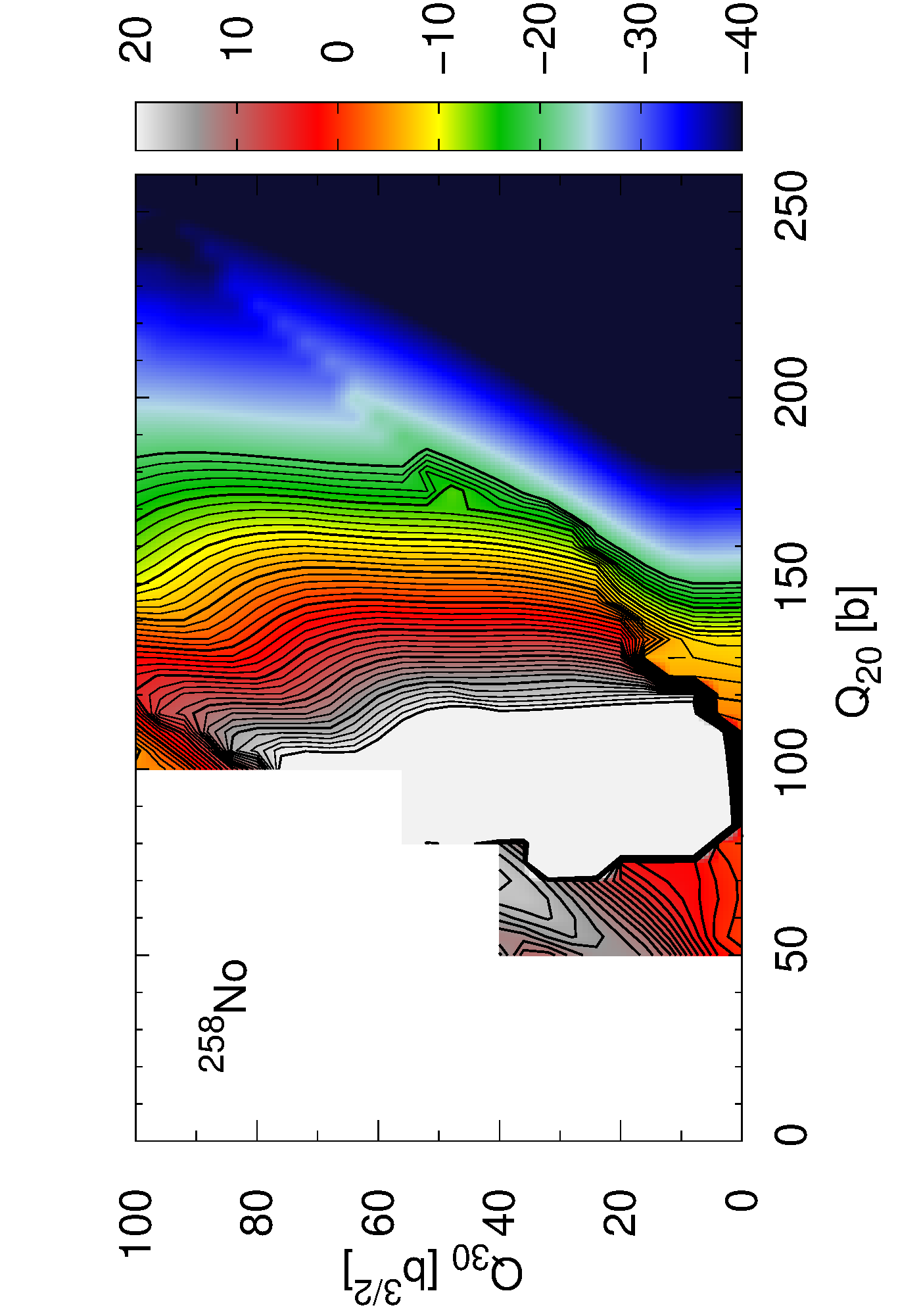}

\caption{The same as in Fig.~\ref{PEScfno} but for the fusion valleys are 
plotted. \label{PESfusion}}

\end{figure}

We have already shown that the description of fission in terms of a  one dimensional or even a
two-dimensional PES may lead to many misunderstandings. Two or more 
different configurations can be obtained in the same location of the PES in those
cases. This problem has been observed 
while discussing the region between the compact symmetric and the 
asymmetric valleys. Similar issues can arise if one analyses the 
so-called fusion channel with two separate fragments. Decreasing the 
quadrupole moment of the system in this region leads to approaching 
fragments which are much closer than what can be achieved in the scission 
configuration. In consequence, the fusion valley covers a much larger 
area than presented in Fig. \ref{PEScfno}. This surface, mostly 
"hidden" under the asymmetric fission valley from Fig \ref{PEScfno}, is 
shown in Fig. \ref{PESfusion}. Only in the lowest 
quadrupole deformation region, the fragments are close to each other, and 
Coulomb energy is so high that the fusion valley climbs up above the 
fission one. Since the global energy minimum for a given quadrupole 
moment may be in the post-scission configuration, an important physical 
problem appears. Should we take this solution as part of the PES 
leading to fission or rather keep staying in the fission valley as long 
as it is possible.

Two alternative approaches can be found in Fig. 1 of Ref. 
\cite{Regnier19} by Regnier {\it et al.} and Fig. 1 of 
Ref. \cite{Warda_2015} by Warda {\it et al.} Both Figures show the PES 
of $^{258}$Fm calculated with the same model and interaction 
(HFB theory and Gogny D1S force). 
Nevertheless, the shape of the asymmetric fission valley is different. 
In the first plot, it is narrow and equipotential (the lines are mainly 
horizontal), whereas in the second case, the asymmetric valley is wide 
and the lines of constant energy are rather vertical. The source of 
the difference comes from the distinct strategy of selecting the local 
minima for the surface. Regnier {\it et al.} selected the lowest of the 
local minima for the given constraints.  Warda {\it et al.} put more 
attention to the continuity of the changes in the shape and 
preservation of the valley in which a nucleus was in the previous phase 
of evolution. In the first approach, the scission line is localized at 
a much lower elongation, and the PES includes a larger part of the 
fusion valley.

We would like to stress that, unexpectedly,  both figures give the 
correct surface in two dimensions showing the importance of considering 
multidimensional PESs. Restricting the deformation space to just one, 
two or three dimensions simplifies the interpretation of the results as 
well as its graphical representation, but it always hides information 
and can veil our understanding of the nature of the fission process. 

We would like to point out that one can observe an internal structure 
of the PES in the fusion channel. The surface is usually smooth with 
mass and deformations of the fragments gradually evolving with octupole 
moment. A rapid drop in the energy indicates an abrupt change of the 
fragment configuration.

\begin{figure}
\includegraphics[angle=0, width=0.9\columnwidth]{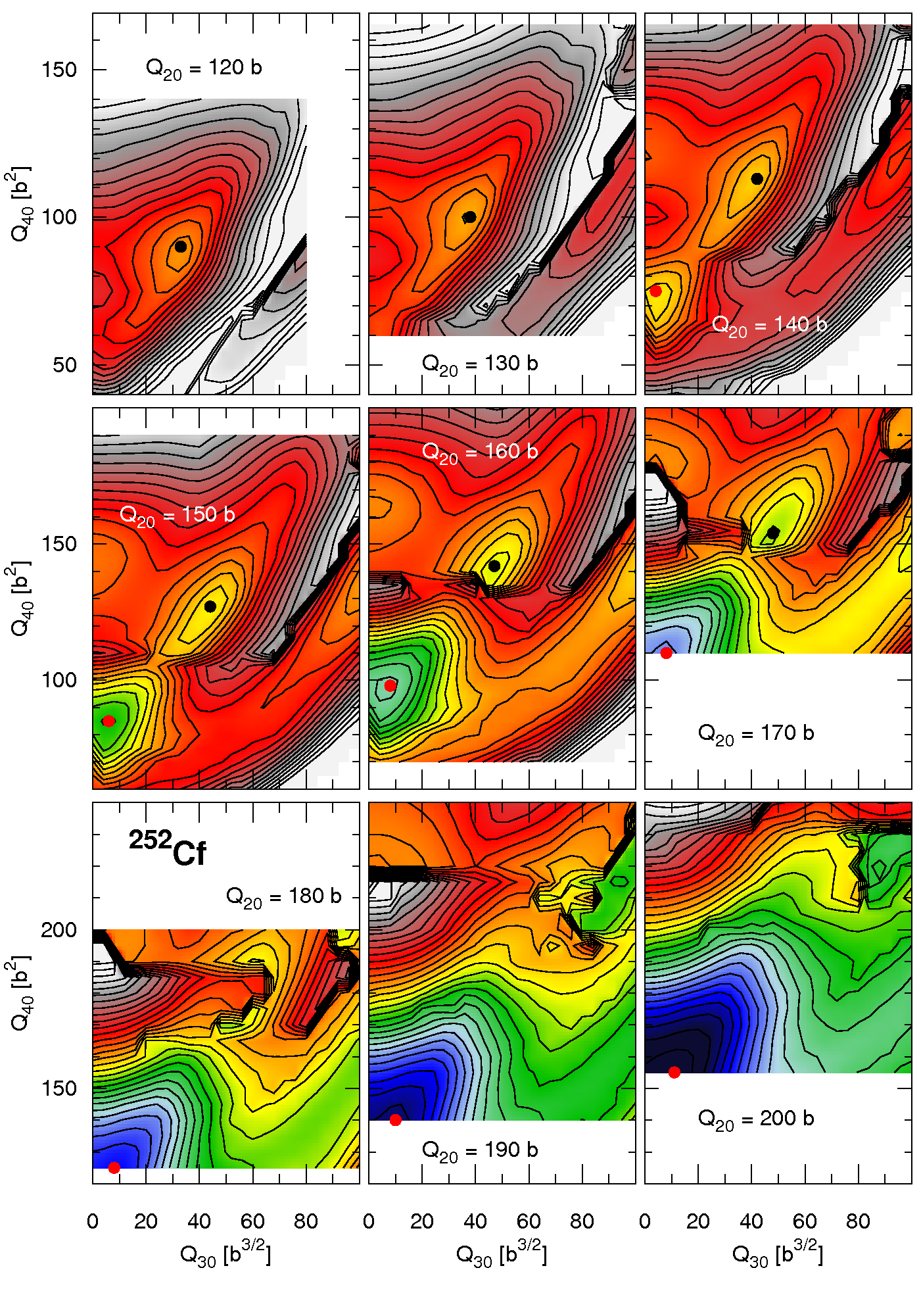}\\
\includegraphics[angle=0, width=0.9\columnwidth]{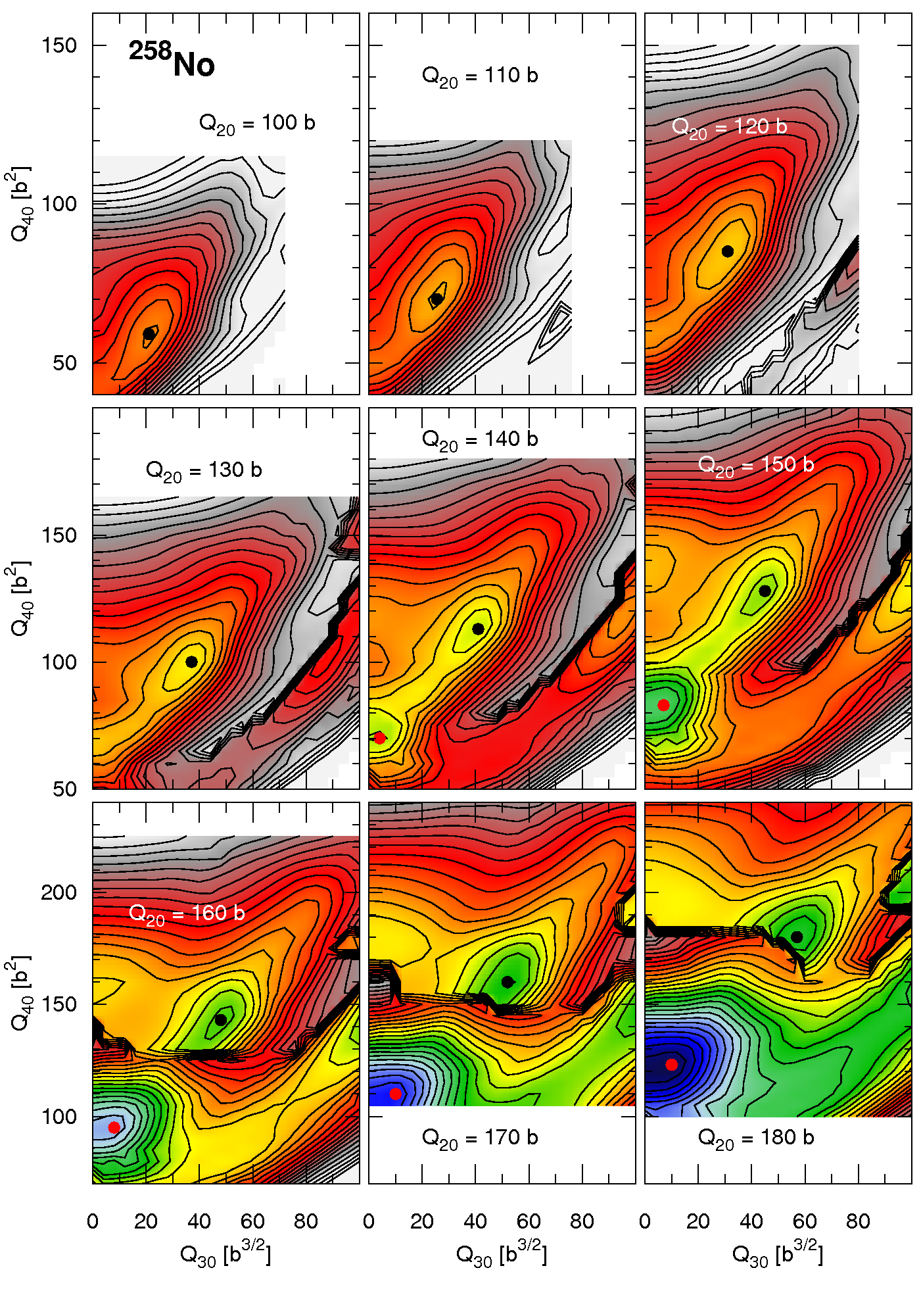}

\caption{The same as in Fig.~\ref{PESfusion} but for cross-sections in 
$Q_{30} - Q_{40}$ plane. \label{PESfusionQ4}}

\end{figure}

The extended fusion valley can also be seen in Fig. \ref{PESfusionQ4} 
where, for fixed quadrupole moments, the PES as a function of the 
octupole and hexadecapole moments are plotted. Again we can see the 
fusion valley even at low quadrupole moments. It extends towards much 
larger hexadecapole moments than it was shown in Figs. \ref{q4_cf} and 
\ref{q4_no}. This analysis provides one more, not very optimistic, 
conclusion. Applying a triple constraint (on quadrupole, octupole and 
hexadecapole moments) and preparing a three-dimensional surface in the 
self-consistent calculation does not guarantee the uniqueness of the 
solution. We still can obtain completely distinct shapes of the density 
distribution depending on the initial configuration.

\section{Conclusions}

We have investigated the fission barriers of two heavy actinide 
nuclei: $^{252}$Cf and $^{258}$No using the self-consistent 
microscopic approach. The calculations were made using multiple 
constraints on the quadrupole, octupole, and hexadecapole moments. A 
detailed analysis has shown a complicated structure of the potential 
energy surface at large quadrupole deformations of the nucleus. The 
competition between local minima at a given quadrupole moment: compact, 
elongated and symmetric elongated determines the fission mode and the 
experimentally observed fragment mass distribution. We have shown that 
the scission may occur at a quadrupole deformation smaller than 
the one at the end of the fission path in the minimum of the valley on 
the potential energy surface.

The calculations using three constraints give a much more 
complete description of 
the potential energy surface. Nevertheless, it is possible to find 
distinct solutions corresponding to the same values of the quadrupole, 
octupole, and hexadecapole moments. These configurations create various 
layers of the potential energy surface in the same place of two- or 
three-dimensional maps. It makes the description of the 
fission even more complicated.

Reducing the full space of deformation to two-dimensions creates a 
potential energy surface that is not continuous. Rapid changes of the 
configuration of density distribution may be found even at seemingly 
smooth surfaces. We therefore conclude that the jumps between surfaces 
must be discussed in the analysis of the fission process.

The present analysis is based on the HFB approximation, but it also 
applies to any constrained, self-consistent type of calculations. 
Problems with multiple minima and surface discontinuities also affect 
macroscopic-microscopic models, although it is much easier to control 
the whole spectrum of nuclear shapes there.

The discussion of the determination of the potential energy surface 
is not the only relevant issue in the analysis of fission. Many other aspects
also have a significant impact on the dynamics of this process. We should mention 
here the influence of the inertia parameter and pairing degrees of 
freedom that shall also be investigated in the future.

\acknowledgments

M.W. acknowledges support by the Polish National Science Center under 
Contract No. 2018/30/Q/ST2/00185. The  work of LMR is supported by 
Spanish Ministry of Economy and Competitiveness (MINECO) Grant No. 
PGC2018-094583-B-I00.







\bibliographystyle{apsrev}
\bibliography{pes}


\end{document}